\documentclass[11pt]{article}
\usepackage{amsmath,amsthm,latexsym,amssymb,amsfonts,epsfig,psfrag}

\makeatletter \@addtoreset{equation}{section} \makeatother

\def\be{\begin{equation}}
\def\ee{\end{equation}}
\def\ba{\begin{eqnarray}}
\def\ea{\end{eqnarray}}

\newcommand\nn{\nonumber}
\newcommand\q{\quad}
\def\Nl{{\mathchoice
{\setbox0=\hbox{$\displaystyle\rm N$}\hbox{\hbox to0pt
{\kern0.4\wd0\vrule height0.9\ht0\hss}\box0}}
{\setbox0=\hbox{$\textstyle\rm N$}\hbox{\hbox to0pt
{\kern0.4\wd0\vrule height0.9\ht0\hss}\box0}}
{\setbox0=\hbox{$\scriptstyle\rm N$}\hbox{\hbox to0pt
{\kern0.4\wd0\vrule height0.9\ht0\hss}\box0}}
{\setbox0=\hbox{$\scriptscriptstyle\rm N$}\hbox{\hbox to0pt
{\kern0.4\wd0\vrule height0.9\ht0\hss}\box0}}}}
\def\Zl{{\mathchoice
{\setbox0=\hbox{$\displaystyle\rm Z$}\hbox{\hbox to0pt
{\kern0.4\wd0\vrule height0.9\ht0\hss}\box0}}
{\setbox0=\hbox{$\textstyle\rm Z$}\hbox{\hbox to0pt
{\kern0.4\wd0\vrule height0.9\ht0\hss}\box0}}
{\setbox0=\hbox{$\scriptstyle\rm Z$}\hbox{\hbox to0pt
{\kern0.4\wd0\vrule height0.9\ht0\hss}\box0}}
{\setbox0=\hbox{$\scriptscriptstyle\rm Z$}\hbox{\hbox to0pt
{\kern0.4\wd0\vrule height0.9\ht0\hss}\box0}}}}
\def\Ql{{\mathchoice
{\setbox0=\hbox{$\displaystyle\rm Q$}\hbox{\hbox to0pt
{\kern0.4\wd0\vrule height0.9\ht0\hss}\box0}}
{\setbox0=\hbox{$\textstyle\rm Q$}\hbox{\hbox to0pt
{\kern0.4\wd0\vrule height0.9\ht0\hss}\box0}}
{\setbox0=\hbox{$\scriptstyle\rm Q$}\hbox{\hbox to0pt
{\kern0.4\wd0\vrule height0.9\ht0\hss}\box0}}
{\setbox0=\hbox{$\scriptscriptstyle\rm Q$}\hbox{\hbox to0pt
{\kern0.4\wd0\vrule height0.9\ht0\hss}\box0}}}}
\def\Rl{{\mathchoice
{\setbox0=\hbox{$\displaystyle\rm R$}\hbox{\hbox to0pt
{\kern0.4\wd0\vrule height0.9\ht0\hss}\box0}}
{\setbox0=\hbox{$\textstyle\rm R$}\hbox{\hbox to0pt
{\kern0.4\wd0\vrule height0.9\ht0\hss}\box0}}
{\setbox0=\hbox{$\scriptstyle\rm R$}\hbox{\hbox to0pt
{\kern0.4\wd0\vrule height0.9\ht0\hss}\box0}}
{\setbox0=\hbox{$\scriptscriptstyle\rm R$}\hbox{\hbox to0pt
{\kern0.4\wd0\vrule height0.9\ht0\hss}\box0}}}}
\def\Cl{{\mathchoice
{\setbox0=\hbox{$\displaystyle\rm C$}\hbox{\hbox to0pt
{\kern0.4\wd0\vrule height0.9\ht0\hss}\box0}}
{\setbox0=\hbox{$\textstyle\rm C$}\hbox{\hbox to0pt
{\kern0.4\wd0\vrule height0.9\ht0\hss}\box0}}
{\setbox0=\hbox{$\scriptstyle\rm C$}\hbox{\hbox to0pt
{\kern0.4\wd0\vrule height0.9\ht0\hss}\box0}}
{\setbox0=\hbox{$\scriptscriptstyle\rm C$}\hbox{\hbox to0pt
{\kern0.4\wd0\vrule height0.9\ht0\hss}\box0}}}}
\def\Hl{{\mathchoice
{\setbox0=\hbox{$\displaystyle\rm H$}\hbox{\hbox to0pt
{\kern0.4\wd0\vrule height0.9\ht0\hss}\box0}}
{\setbox0=\hbox{$\textstyle\rm H$}\hbox{\hbox to0pt
{\kern0.4\wd0\vrule height0.9\ht0\hss}\box0}}
{\setbox0=\hbox{$\scriptstyle\rm H$}\hbox{\hbox to0pt
{\kern0.4\wd0\vrule height0.9\ht0\hss}\box0}}
{\setbox0=\hbox{$\scriptscriptstyle\rm H$}\hbox{\hbox to0pt
{\kern0.4\wd0\vrule height0.9\ht0\hss}\box0}}}}
\def\Ol{{\mathchoice
{\setbox0=\hbox{$\displaystyle\rm O$}\hbox{\hbox to0pt
{\kern0.4\wd0\vrule height0.9\ht0\hss}\box0}}
{\setbox0=\hbox{$\textstyle\rm O$}\hbox{\hbox to0pt
{\kern0.4\wd0\vrule height0.9\ht0\hss}\box0}}
{\setbox0=\hbox{$\scriptstyle\rm O$}\hbox{\hbox to0pt
{\kern0.4\wd0\vrule height0.9\ht0\hss}\box0}}
{\setbox0=\hbox{$\scriptscriptstyle\rm O$}\hbox{\hbox to0pt
{\kern0.4\wd0\vrule height0.9\ht0\hss}\box0}}}}
\newcommand{\bd}{\mathbf d}

\title{Coarse graining free theories with gauge symmetries:\\ the linearized case}
\author{Benjamin Bahr$^{1,2}$, Bianca Dittrich$^1$, Song He$^1$\\
\small   $^1$ MPI f. Gravitational Physics, Albert Einstein Institute,\\
 \small Am M\"uhlenberg 1, D-14476 Potsdam, Germany \\
\small $^2$ DAMTP, University of Cambridge,\\
\small  Wilberforce Road, Cambridge CB3 0WA, UK }

\date{}

\begin{document}

\maketitle

\begin{abstract}
\noindent
Discretizations of continuum theories often do not preserve the gauge symmetry content. This occurs in particular for diffeomorphism symmetry in general relativity, which leads to severe difficulties both in canonical and covariant quantization approaches. We discuss here the method of perfect actions, which attempts to restore gauge symmetries by mirroring exactly continuum physics on a lattice via a coarse graining process. Analytical results can only be obtained via a perturbative approach, for which we consider the first steps, namely the coarse graining of the linearized theory. The linearized gauge symmetries are exact also in the discretized theory, hence we develop a formalism to deal with gauge systems. Finally we provide a discretization of linearized gravity as well as a coarse graining map and show that with this choice the 3D linearized gravity action is invariant under coarse graining.
\end{abstract}

\tableofcontents

\section{Introduction}

Discretizations of field theories have  become a viable tool for both classical and quantum physics. On the one hand, numerical treatments of for instance gravity require discretization, on the other hand, lattice quantum field theories give access to non--perturbative physics. One might even expect that discrete structures will play a fundamental  role in quantum gravity -- as opposed to just providing an auxiliary  UV cut--off. Indeed in many approaches to quantum gravity  such discrete structures either appear as fundamental ingredients, as derived from a continuum quantization or as auxiliary structures.

Independent from the interpretation of these discrete structures as fundamental or auxiliary, the question arises how to retrieve the  continuum physics we experience at larger scales from the microscopic models involving discrete structures.  This applies in particular to the emergence of continuum symmetries, as these influence physical predictions as well as the interpretation of the models.

For general relativity diffeomorphism symmetry plays an extraordinary important role as it is deeply intertwined with the dynamics of the theory. Unfortunately diffeomorphism symmetry is usually broken by discretization \cite{lollreview,Loll:1997iw,Dittrich:2008pw, bahrdittrich1,Dittrich:2009fb}. This leads to severe difficulties both for covariant and canonical quantization approaches.

 In the covariant approach, breaking of gauge symmetries leads to additional degrees of freedom. The gauge modes which for an exact symmetry completely decouple from the physical modes will become relevant and couple to the remaining modes if the gauge symmetries are broken. Hence these modes have to be taken into account in a quantization (and cannot be gauge fixed), but should become irrelevant in the continuum limit.

In the canonical approach, the dynamics of general relativity is encoded in the Hamiltonian and diffeomorphism constraints. The central problem in the canonical quantization program is to implement these constraints into the quantization. Here,  breaking of diffeomorphism symmetry leads to inconsistencies in the dynamics defined by these constraints, which severely impedes the quantization of the constrained theory. This has been a huge obstacle to canonical gravity lattice models \cite{lollreview}, see however the suggestions in \cite{Thiemann:2003zv, Gambini:2005jm}. For further discussions of these points and related issues see \cite{lollreview, Dittrich:2008pw, bahrdittrich1, Dittrich:2009fb, wp, friedman, miller, morse, hwgauge, menotti} and references therein.

Discretizations that would preserve some notion of diffeomorphism symmetry\footnote{We will discuss in section \ref{Sec:LinearizedGravity} what kind of diffeomorphism symmetry one would expect in a discretized theory.} would therefore be very appreciated. Indeed there is an approach to construct such discretizations. The associated discrete actions, encoding the discretized dynamics, are called perfect actions \cite{Hasenfratz:1997ft, Bahr:2009qc}. The basic idea is to map continuum physics onto the lattice by a coarse graining process. The resulting lattice theory will then mirror exactly continuum dynamics in its coarse grained observables. Hence one would also expect the continuum symmetries to be present in this lattice dynamics (at least those, which have not been absorbed by the coarse graining process).

Furthermore this process might lead to `lattice independent' lattice theories. That is not only are lattice artifacts avoided and predictions should not depend on the choice of lattice. More fundamentally, observables of the theory should not depend in any way on lattice sites. This corresponds to the requirement of diffeomorphism invariant -- hence coordinate independent -- observables in the continuum. Such observables are known from topological lattice models, where there are only finitely many global observables. For 4D gravity, we however expect the number of such observables to scale with the number of lattice sites.

The coarse graining process can be performed in two different ways. One is to consider a `block transformation from the continuum', i.e. to have only one coarse graining step from an infinitesimal lattice constant to a finite one \cite{Bietenholz:1999kr}. This should result immediately into a theory where continuum symmetries are preserved. There is a  disadvantage however, namely that coarse graining always involves solving at least partially the theory. Hence this method requires some control over the solutions. Coarse graining can be also performed in many small steps from smaller to larger scales. This usually allows one to introduce approximations, in the (frequent) case, that exact evaluations are not possible and is related to the ideas of Wilsonian renormalization group flow \cite{Wilson:1973jj}.

As many quantum gravity models are discrete on the microscopic scale one would here rather adopt the second strategy. Indeed, even classical gravity being a very hard to solve theory, we cannot expect to obtain a perfect action easily. The prospect is rather to understand the coarse graining process better and to derive conditions on the microscopic theory, so that diffeomorphism symmetry will arise at macroscopic scales.

Mostly, discrete gravity models can only be solved numerically. An alternative is to consider a perturbative approach and to coarse grain the theory order by order in the perturbations. In this work we will start with  gravity linearized on a flat background. In this case the (linearized) gauge symmetries of the continuum are typically still realized. These are however broken in the higher order theory. This actually leads to inconsistencies in the perturbative approach \cite{Dittrich:2009fb}. From the higher order equation of motions  non-linear consistency equations arise, which determine  the gauge degrees of freedom, present at lower order.

To avoid these inconsistency issues one has to `perfectionize' the action order by order. That is a second order perfect action will allow a consistent solution of the theory truncated at third order -- where all the lower order gauge parameters remain free.

We therefore consider in this work the first step -- the coarse graining of quadratic actions, that is free theories. Here the main point we address, is to coarse grain theories with gauge symmetries in a gauge covariant way, that is without involving a gauge fixing.  This is inevitable for discrete gravity, as due to the breaking of diffeomorphism symmetries gauge fixing is not a viable procedure anymore starting with the third order theory. We will leave the consideration of higher order perturbations  for future work.

As will be explained in section \ref{Sec:LinearizedGravity} the expected gauge symmetries for discrete gravity have a very geometric interpretation. We will therefore always be motivated in our choice of discrete action and the coarse graining procedure by geometric reasoning. For (linearized) discrete gravity we will derive an action and a coarse graining map derived from Regge gravity \cite{Regge}, which in itself is a very geometric discretization of gravity. An indication that this choice is suitable, will be provided by 3D gravity, whose (linearized) discrete action will turn out to be form invariant under coarse graining, as expected from a topological theory.

The plan of the paper is as follows.
We will start by reviewing basic coarse graining for free fields in section \ref{conv}, also to introduce our conventions and notations. This formalism will be applied to coarse grain the free scalar field both from a finite lattice to a coarse grained finite lattice and by coarse graining from the continuum to a finite lattice, section \ref{Sec:ScalarField}.

We will then develop the formalism in order to coarse grain theories with gauge degrees of freedom in section \ref{Sec:CoarseGrainingWithGauge}. To this end we will first discuss the behavior of gauge symmetries under coarse graining in section \ref{Sec:SubGaugeDOFUnderCoarse}. In particular we will argue, which kind of discrete diffeomorphism symmetry one might expect for a perfect action for discrete gravity. The formalism will then be applied to electromagnetism in section \ref{Sec:Electromagnetism} and we will show that for 2D electromagnetism the action is form invariant under coarse graining.

Finally we will provide a discrete action and a coarse graining map for linearized gravity in section \ref{Sec:LinearizedGravity}. Again we will show that with this choice the 3D linearized gravity action is form invariant under coarse graining. We will end with a short summary and an outlook.

The appendix \ref{appmetric} contains some material on the relation between Regge calculus and the discretization for gravity employed here, so that we can provide a geometric derivation of the coarse graining map. In appendix \ref{somesums} we evaluate some sums, which are needed for the coarse graining of 2D electromagnetism and 3D gravity.

\section{Coarse graining of free fields}\label{conv}

Here we will consider the coarse graining of free field theories without gauge symmetries on the lattice. We derive a general formula for the coarse grained action that we will apply to a free scalar field, reproducing the results of \cite{Bell:1974vv}.

We will consider fields $\phi_A$, with $A$ a yet to be specified index, on a $d$-dimensional periodic lattice with $N^d$ sites $x=(0,\ldots,0),\ldots (N-1,\ldots, N-1)$. For free fields the action will be a quadratic functional of the fields of the form
 \ba\label{b1}
S=\tfrac{1}{2}\sum_{A,B}\sum_{x,y}   \phi_A(x)\, m_{AB} (x-y) \,  \phi_B(y)   \q .
\ea
Here we assume that $m_{AB}$ does only depend on the difference $(x-y)$, that is that the action is invariant under (lattice) translations.  For most of the discussion we will work with the Fourier transformed fields. All the general formulas can easily adapted back for fields $\phi_A(x)$ in real space. We will use Fourier transformed fields here in order to apply the formulas at once to concrete examples.

Introducing the momentum labels  $p=(0,\ldots,0),\ldots (N-1,\ldots N-1)$ we define
\ba\label{b2}
\phi_A(p)=\sum_x e^{-2\pi i \frac{p\cdot x}{N}}\, \phi_A(x) \q ,\q\q  \phi_A(x)=\frac{1}{N^d}\sum_p e^{2\pi i \frac{p\cdot x}{N}}\, \phi_A(p) \q .
\ea
For the inverse we used that the delta--function on the ($N$-periodic) lattice is given by
\ba\label{b4}
\delta^{(N)}(p) =\frac{1}{N^d}\sum_x e^{2\pi i \frac{p\cdot x}{N}}  \q .
\ea
The action in the Fourier transformed fields is then
\ba\label{b5}
S&=&
 \frac{1}{2 \,N^d}    \sum_{A,B,p}    \phi_A(p)  m_{AB}(p) \phi_B(-p)  \q\q \text{with}\q\q m_{AB}(p)\;=\;\sum_x e^{2\pi i \frac{p\cdot x}{N}}  \,m_{AB}(x)   \, .\q
\ea

The action (\ref{b1}) will be varied under the conditions that the field values $\phi(x)$ sum up to the coarse grained fields $\Phi(X)$ on a coarse grained lattice with sites $X=(0,\ldots,0),\ldots (N'-1,\ldots,N'-1)$ where
$N=L \, N' $.
The extrema (or solutions) of the action (\ref{b1})  obtained with these conditions will be functions of the coarse grained fields $\Phi$. Reinserting these solutions into the action (\ref{b1}) we obtain a coarse grained action $S'$ as a function of the coarse grained fields $\Phi$:
\ba\label{defcg}
S'\left[\Phi\right]= \underset{ \phi,  \,B \phi=\Phi}{\text{extr}} \, S
\ea
where $B$ is the coarse graining map.
Varying this new action $S'$ with respect to the fields $\Phi$ we will find new solutions $\Phi_s$ describing the dynamics of the theory on the coarse grained lattice. The solutions $\Phi_s$ encode however the dynamics of the original  lattice, as these solutions can be obtained by coarse graining the solutions of the action S (without adding any conditions on the fields $\phi$). Namely, what has been done, is to split the variational problem for the action $S$ into two parts: first one looks for extrema under the condition that the $\phi$ coarse grain to $\Phi$. Then one varies the conditions $\Phi$, so that one re--obtains the extrema of the action $S$
\ba
  \underset{\Phi}{\text{extr}}  \,  \underset{ \phi,  \,B \phi=\Phi}{\text{extr}} \, S &=& \underset{\phi}{\text{extr}} \, S  \q .
\ea
 Just that we have now only access to these extrema via the coarse grained fields $\Phi$.

We will write the  coarse graining map as
\ba\label{b8}
\Phi_A(X)= \sum_{B,x}  B_{AB}(X,x) \,\phi_B (x)
\ea
which for the Fourier transformed fields gives
\ba\label{b9}
\Phi_A(P)
&=& \q \,\,\, \sum_X  e^{-2\pi i \frac{P\cdot X}{N'}}\, \Phi_A(X)
\;=:\; \frac{1}{N^d} \sum_{B,p} B_{AB}(P,p) \, \phi_B(p)  \q
\ea
where
\ba\label{b10}
 B_{AB}(P,p)&=&\sum_{X,x} e^{-2\pi i \frac{P\cdot X}{N'}} B_{AB}(X,x) e^{2\pi i \frac{p\cdot x}{N}}  \q .
\ea

The coarse graining conditions (\ref{b9}) can be added to the action (\ref{b5}) with Lagrange multipliers $\lambda(-P)$, so that we have to vary
\ba\label{b11}
S_\lambda=   \frac{1}{2 N^d}    \sum_{A,B,p}    \phi_A(p)  m_{AB}(p) \phi_B(-p) +   \sum_{A,P} \lambda_A(-P)
\bigg( \Phi_A(P)-\frac{1}{N^d}\sum_{B,p} B_{AB}  \, \phi_B(p) \bigg)    \, .\q\q
\ea

This gives rise to the equations of motion
\ba\label{b12}
m_{AB}(p) \,\phi_B(-p) \;=\; \sum_{C,P}  \lambda_C(-P)\, B_{CA}(P,p)   \, ,\q\q
\label{b13}
\Phi_A(P)\;=\; \frac{1}{N^d}\sum_{B,p}  B_{AB}(P,p)\,\phi_B(p)   \, .\q
\ea
Assuming that $m(p)$ is invertible we can write
\ba\label{b14}
N^d\,\Phi_A (P) &=& \sum_{B,p} B_{AB}(P,p) \phi_B(p)\nn\\
 &=& \sum_{B,C,D,Q,p}       B_{AB}(P,p)  \, \,   (m^{-1})_{BD}(-p) \,  \,   B_{CD}(-Q,-p)  \, \,\,   \lambda_C(Q)  \, .\q
\ea
On the other hand we can rewrite the action with the help of (\ref{b12}) to
\ba\label{b15}
S&=&\frac{1}{2N^d}   \sum_{A,B,p}    \phi_A(p)\,  m_{AB}(p) \phi_B(-p) \nn\\
&=&   \frac{1}{2N^d}   \sum_{A,C,P,p}    \phi_A(p)  \,  \lambda_C(-P)\, B_{CA}(P,p)  \nn\\
&=& \frac{1}{2}\sum_{A,P} \Phi_A(P)\,\lambda_A(-P)   \q .
\ea
Hence, this time assuming that the matrix $B \cdot m^{-1} \cdot B$ appearing in the last line of (\ref{b14}) is invertible, we obtain for the coarse grained action
\ba\label{b16}
S'&=& \frac{N^d}{2} \sum_{A,B,P,Q}   \Phi_A(P) \,\, M_{AB}(-P,-Q)\,\, \Phi_B(-Q)
\ea
where
\ba\label{b17}
(M^{-1})_{AB}(P,Q)&=& \sum_{C,D,p}       B_{AC}(P,p)  \,\,    (m^{-1})_{CD}(-p) \, \,    B_{BD}(-Q,-p)   \q .
\ea

\section{Example: Perfect action for scalar field}\label{Sec:ScalarField}

\subsection{The coarse graining}

Here we will apply the general formalism to a free scalar field discretized on a regular hyper--cubical lattice.
We adopt the following (Wick rotated) action for a free scalar field $\phi$ on a $d$-dimensional periodic lattice
\ba\label{b18}
S&=&\frac{1}{2} \sum_{x,y}  \phi(x) \,  m(x-y)\, \phi(y)  \nn\\
&=&\frac{a^d}{2} \sum_{x,y}   \phi(x) \, \bigg( \Delta(x,y) +\mu^2 \delta^{(N)}(x,y) \bigg)\, \phi(y)  \q .
\ea
The Laplace operator on the lattice is defined as
\ba\label{b19}
 \Delta(x,y)&=&\frac{1}{a^2}  \sum_b \bigg(2 \delta^{(N)}(x,y) - \delta^{(N)}(x,y+e_b) -\delta^{(N)}(x,y-e_b)\bigg)
\ea
with $a \propto 1/N$ the lattice constant and $e_b$ the lattice vectors in direction $b=1,\ldots,d$.
Its Fourier transformation is
\ba\label{b20}
\Delta(p)
&=& \frac{1}{a^2} \sum_b \bigg(2- e^{2\pi i \frac{p_b}{N} } -e^{-2\pi i \frac{p_b}{N} }  \bigg)
\;\; =:\;\; \frac{1}{a^2} \sum_b  k_b \bar{k}_b
\ea
where we defined $k_b=(1-e^{2\pi i \frac{p_b}{N} } )$ and $\bar{k}_b=   (1-e^{-2\pi i \frac{p_b}{N} } )$.
For the Fourier transformed action we obtain
\ba\label{b21}
S=\frac{a^d}{2 N^d} \sum_p \phi(p) \big(\Delta(p)+\mu^2) \phi(-p) \q .
\ea

The coarse grained scalar field $\Phi(X)$ will be defined as the sum over the fields $\phi(x)$ over all lattice sites $x$ in a box associated to $X$.
\ba\label{b22}
\Phi(X)
        &=& \sum _x B(X,x) \, \, \phi(x)
       \;\;:=\;\;  \sum_x  \, b \sum_z  \delta^{(N)}(x, LX+z) \,\,\phi(x)
\ea
where $b$ is some rescaling constant and $z$ assumes the values $z=(0,\ldots,0),\ldots, (L-1,\ldots,L-1)$. (Remember that $N=LN'$.)  Fourier transforming  the matrix $B$ gives
\ba\label{b23}
B(P,p)\;:=\; \sum_{X,x}      e^{-2\pi i \frac{P\cdot X}{N'} }  B(X,x)   e^{2\pi i \frac{p\cdot x}{N} }
&=& \sum_{X,x}    \,b \sum_z     e^{-2\pi i \frac{P\cdot X}{N'} } \, \,\delta^{(N)}(x, LX+z)   \,\,e^{2\pi i \frac{p\cdot x}{N} } \q\q \nn\\
&=& b\, N'^d \, \, \delta^{(N')}(P-p)\, \sum_z  e^{2\pi i \frac{p\cdot z}{N} }  \nn\\
&=&  b\, N'^d\,  \, \delta^{(N')}(P-p)\, \, \prod_a \frac{1-e^{2\pi i \frac{L\,p_a}{N} }} {1-e^{2\pi i \frac{p_a}{N} }} \nn\\
&=&  b\, N'^d  \,\, \delta^{(N')}(P-p)\,\,  \prod_a  \frac{K_a}{k_a} \q . \q\q\q
\ea
where for the sum over $z=(0,\ldots,0),\ldots,(L-1,\ldots,L-1)$ we used that it is a product of geometric series. In the last line we introduced $K_b=1-e^{2\pi i \frac{P_b}{N'} }$. Later we will also abbreviate $\bar{K}_b=1-e^{-2\pi i \frac{P_b}{N'} }$.

Now we already have all the prerequisites to apply formula (\ref{b16},\ref{b17}) for the coarse grained action:
\ba\label{b24}
M^{-1}(P,Q) &=&\sum_p B(P,p)\,\, m^{-1}(-p)\, \, B(-Q,-p) \nn\\
&=& b^2\,N'^{2d} \sum_p  \delta^{(N')} (P-p) \,\,\delta^{(N')}(Q-p) \,\,\frac{a^{2-d}}{\sum_b k_b\bar{k}_b+a^2 \mu^2} \,\,\prod_b \frac{K_b \bar{K}_b}{k_b \bar{k}_b} \nn\\
&=&
 a^{2-d}b^2\,N'^{2d}  \delta^{(N')}(P-Q)\,\sum_r \left( \frac{1}{\sum_b k_b\bar{k}_b+a^2 \mu^2} \,\,\prod_b \frac{K_b \bar{K}_b}{k_b \bar{k}_b}\right)_{| p=P+N'r} \; .\q\q
\ea
In the last line the sum is over $r=(0,\ldots,0),\ldots,(L-1,\ldots,L-1)$ and the $k_b$ depend via $k_b=1-e^{2\pi i \frac{p_b}{N} }$ on $p$ (whereas the $K_b$ depend only on $P$).  The coarse grained action is then given by
\ba\label{b25}
S'=\frac{1}{2}\,\frac{L^d \,a^{d-2}}{ N'^d\, b^2} \,  \sum_P  \Phi(P)    M'(P) \Phi(-P)
\ea
where
\ba\label{b26}
M'(P)= \left(
\sum_r \left( \frac{1}{\sum_b k_b\bar{k}_b+a^2 \mu^2} \,\,\prod_b \frac{K_b \bar{K}_b}{k_b \bar{k}_b}\right)_{| p=P+N'r}
\right)^{-1}       \q .
\ea

The sum over $r$ can only be performed analytically for one dimensional systems, $d=1$, see below. Also the action (\ref{b25}) (Fourier transformed back to $X$ labels) will in general be non--local\footnote{These couplings are however exponentially decaying with distance and sometimes such a case is still referred to as a theory with local couplings. We termed it non--local, to distinguish it from the lattice actions, which are mostly used in discrete gravity approaches, such as Regge gravity and spin foam models. There the couplings are restricted to nearest neighbours.}, that is involve couplings between non-neighbouring lattice sites.


\subsection{Blocking from the continuum}\label{blockcont}

To obtain the action coarse-grained from the continuum, we can
iterate the blocking procedure infinite times for finite $L$ to obtain a fixed point. Alternatively \cite{Bietenholz:1999kr}, we
can directly `block from the continuum', to obtain an action on a finite lattice, mirroring the continuum theory.

We will consider $T$--periodic continuum fields $\phi(\chi)$, with $\chi\in [0,T)^d$. For the Fourier transformation we adopt the conventions
\ba\label{bia02}
\phi(\kappa)\;=\;\int_{[0,T)^d} \bd^d \chi \,\,e^{-2\pi i \frac{\kappa \cdot \chi}{T}} \,\, \phi(\chi)\q,\q\q \phi(\kappa)=\frac{1}{T^d} \sum_{\kappa \in \mathbb{Z}^d} e^{2\pi i \frac{\kappa \cdot \chi}{T}} \,\,\phi(\kappa)  \, ,
\ea
so that the momentum label $\kappa$ takes values in $\mathbb{Z}^d$. The continuum action is given by
\ba\label{bia03}
S_c \,=\, \frac{1}{2}\int_{[0,T)^d} \bd^d \chi\,\, \phi(\chi)\,\left( -\partial^2_\chi + \mu_c^2 \right) \phi(\chi) \;=\;
\frac{1}{2T^d}\sum_\kappa  \phi(\kappa) \left( \sum_a \left(\tfrac{2\pi}{T} \kappa_a\right)^2 + \mu_c^2\right) \phi(-\kappa) \, . \q\q
\ea
We coarse grain the field by averaging it over cubes of volume $a'^d$, where $a'$ is the lattice constant of the coarse grained lattice, so that $T=N'a'$. That is,
\ba\label{bia03b}
\Phi(X)=b_c\int_{[0,a')^d} \bd^d \epsilon \,\,  \phi(a'X+\epsilon)
\ea
which for the Fourier transformed fields gives
\ba\label{bia04}
\Phi(P):=\sum_{X_c=0}^{N'-1} e^{-2\pi i\frac{P\cdot X}{N'}} \Phi(X)
&=& b_cN'^d \sum_{\kappa \in \mathbb{Z}^d}  \,\delta^{(N')}(P-\kappa)\, \phi(\kappa)\,\prod_{c} \frac{K_c}{\frac{2\pi}{i}\kappa_c}  \; . \q\q\q
\ea
where $K_c=1-e^{2\pi i\frac{P_c}{N'}}$. A derivation completely analogous to the one in section \ref{conv} leads to the coarse grained `perfect' action
\ba\label{bia05}
S'=\frac{1}{2} \frac{1}{b_c^2 T^{d+2} N'^{2d}}\sum_P \Phi(P) M'(P)\Phi(-P)
\ea
where
\ba\label{bia06}
M'^{-1}(P)&=&\sum_{r \in \mathbb{Z}^d} \left( \frac{1}{\sum_b (2\pi\kappa_b )(2\pi \kappa_b )+(T\mu_c)^2} \prod_a \frac{K_a \bar K_a}{ (2\pi \kappa_a)(2\pi\kappa_a)} \right)_{\big{|} \kappa=P+rN'} \q .
\ea


\subsection{One--dimensional system}

Here we will perform the sum in (\ref{b26}) for a one--dimensional system. This will introduce techniques that will be later useful to evaluate the coarse grained actions for (topological) gauge theories.

Following (\ref{b26}) we have to compute
\ba
M'(P)^{-1}&=&\sum_{r=0}^{L-1}
\frac{1}{  \left(      1- e^{ix + \frac{2\pi i }{L}r}     \right) \left(      1- e^{-ix - \frac{2\pi i }{L}r}  \right)   + m^2  } \,\,\,
\frac{    \left(      1- e^{iLx }     \right) \left(      1- e^{-iLx  }  \right)     }{            \left(      1- e^{ix + \frac{2\pi i }{L}r}     \right) \left(      1- e^{-ix - \frac{2\pi i }{L}r}  \right)     } \q\q\q
%
\ea

\noindent where we have defined $m:= a\mu$ and $x:=\tfrac{2\pi}{N}P$. We introduce a different way of writing the mass $m$ by defining
\begin{eqnarray}\label{Gl:DefinitionOflambda}
\tfrac{1}{2}(e^y+e^{-y})=\cosh\left(y\right)\;:=\;\ 1+ \frac{m^2}{2}  \; .
\end{eqnarray}

\noindent From this it follows that
\begin{eqnarray}
\left(1-e^{ix+ \frac{2\pi i}{L}r}\right)\left(1-e^{-ix-\frac{2\pi i}{L}r}\right)\,+\,m^2\;=\;e^{y}\left(1-e^{i(x+iy)+\frac{2\pi i}{L}r}\right)\left(1-e^{-i(x-iy)-\frac{2\pi i}{L}r}\right)\; .
\end{eqnarray}

\noindent Hence we have to evaluate the sum
\begin{eqnarray}
M'(P)^{-1}=
\sum_{r=0}^{L-1}
\frac{    \left(      1- e^{iLx }     \right) \left(      1- e^{-iLx  }  \right)     e^{-y}  }{                \left(      1- e^{ix + \frac{2\pi i }{L}r}     \right) \left(      1- e^{-ix - \frac{2\pi i }{L}r}  \right)   \left(      1- e^{i(x+iy) + \frac{2\pi i }{L}r}     \right) \left(      1- e^{-i(x-iy) - \frac{2\pi i }{L}r}  \right)     } \, . \q
\end{eqnarray}
The basic idea to perform the summation is to rewrite the factors in the denominator into a geometric series, for instance
\ba
\frac{1}{   \left(      1- e^{ix + \frac{2\pi i }{L}r}     \right) }&=&  \frac{1}{   \left(      1- e^{iLx}   \right) }  \,\, \sum_{j=0}^{L-1} e^{ix + \frac{2\pi i }{L}r}    \q .
\ea
In this way we obtain
\ba\label{bia01}
M'(P)^{-1}
&=&
A\,\, \sum_{r=0}^{L-1}\,\,\sum_{j_1,j_2,j_3,j_4=0}^{L-1}\,\, e^{ ix(j_1-j_2+j_3-j_4) -y(j_3+j_4)   +\frac{2\pi i}{L} r (j_1-j_2+j_3-j_4)  }
\ea
where the prefactor $A$ is given by
\ba
A&=& \frac{e^{-y}}{   \left(      1- e^{iL(x+iy)}   \right)  \left(      1- e^{-iL(x-iy)}   \right) } \,\,=\,\, \frac{e^{-y}e^{Ly}}{K\bar{K} + M^2} \q .
\ea
Here we introduced $K=(1-e^{iLx})=1-e^{\frac{2\pi i}{N'}P}$ and the new mass $M$ by
\begin{eqnarray}\label{Gl:DefinitionOflambda2}
\tfrac{1}{2}(e^{Ly}+e^{-Ly})=\cosh\left(Ly\right)\;:=\;\ 1+ \frac{M^2}{2}  \q .
\end{eqnarray}

\noindent Performing the sum over $r$ in (\ref{bia01}) results in a lattice delta function,
$
\sum_{r=0}^{L-1}e^{\frac{2\pi i}{L}rj}\;=\;L\,\delta^{(L)}(j)
$,

\noindent and so we get
\begin{eqnarray}\label{Gl:Sum}
M'(P)^{-1}\;&=&\;L\,A\,\sum_{j_1,j_2,j_3,j_4=0}^{L-1}     e^{ ix(j_1-j_2+j_3-j_4) -y(j_3+j_4) }      \,\delta^{(L)}(j_1-j_2+j_3-j_4)\,\, . \q
\end{eqnarray}
For the given range for the labels $j_i=0,\ldots,L-1$  there are three types of solutions possible for the $L$--periodic delta function. These result if the argument $a(j):=(j_1-j_2+j_3-j_4)$ assumes the values $a(j)=0,\,a(j)=\pm L$, so that we have to consider
\ba
M'(P)^{-1}\;&=&\;L\,A\, \Bigg[
\sum_{\stackrel{j_1,j_2,j_3,j_4=0}{a(j)=0}}^{L-1}  e^{-y(j_3+j_4)}+ (e^{iLx}+e^{-iLx}) \sum_{\stackrel{j_1,j_2,j_3,j_4=0}{a(j)=L}}^{L-1}  e^{-y(j_3+j_4)} \Bigg]
\ea
where we could summarize the $a(j)=\pm L$ case into one summation due to the $\mathbb{Z}_2$-symmetry in the problem.
We rename $j_3-j_4=:J$ and reorder the sums, by counting the possible configurations with $a(j)=0,\pm L$. For $J=0$, there are $L$ possibilities for $j_1, j_2$ (namely both being equal) such that $a(j)=0$. For  $J>0$ there are $L-J$ possibilities for $a(j)=0$, namely whenever $j_2-j_1=J$, similarly for $J<0$.

To obtain $a(J)=L$ (we do not need to consider $-L$, since this has already been taken care of within the sum), we need $J=j_3-j_4>0$. Then there are $J$ possibilities for $j_1-j_2$ to equal $L-J$, and hence satisfying the condition $a(j)=L$.

We conclude:
\begin{eqnarray*}
M'(P)^{-1}\!\!\!\!\!&=&\!\!\!\!\!LA \Bigg[
\sum_{j=0}^{L-1}Le^{-2yj}+2\sum_{j=0}^{L-1}\sum_{J=1}^{L-1-j}\!\!(L-J)e^{-y(2j+J)} +\left(e^{iLx}+e^{-iLx}\right)\!\!\sum_{j=0}^{L-1}\sum_{J=1}^{L-1-j}\!\! \!\!Je^{-y(2j+J)}
\Bigg]  \q\q\q  \nn\\
\!&=&\!\!\!\!\!LA\sum_{j=0}^{L-1}e^{-2yj}\Bigg[L+2L\sum_{J=1}^{L-1-j}e^{-yJ}\;+\;\left(e^{iLx}+e^{-iLx}-2\right)\sum_{J=1}^{L-1-j}Je^{-y J}\Bigg]   \nn
\end{eqnarray*}

The sums can be performed explicitly in a straightforward manner, which results in
\ba
M'(P)^{-1}&=&
\frac{L}{K\bar{K}+M^2} \frac{1}{(\cosh(y)-1)} \left[ L\left(\cosh(Ly)-1\right) +\frac{1}{2}K\bar{K} \left( L-\frac{\sinh(Ly)}{\sinh(y)} \right)\right] \nn\\
 &=& \frac{L}{K\bar K + M^2} \frac{1}{m^2} \left[ L M^2 + K\bar K \left(L-\frac{M \sqrt{4+M^2}}{m \sqrt{4+m^2}}\right)\right] \q .\q\q\q\q
\ea
We obtain for the coarse grained action
\ba\label{1df}
S'&=& \frac{1}{2}\frac{m^2}{N' L a b^2} \sum_P \Phi(P) \, \frac{K\bar K + M^2}{ K \bar K (1- c) + M^2}\,\Phi(-P)
\ea
where
\ba
c=\frac{1}{L} \frac{M \sqrt{4+M^2}}{m \sqrt{4+m^2}}\q .
\ea
Note that the appearance of the factor $K\bar{K}$ in the denominator in the coarse grained action (\ref{1df}) renders it non--local. This can be avoided (but only for one--dimensional systems) by changing the coarse graining map appropriately \cite{Bietenholz:1999kr}. For instance coarse graining by decimation, where the coarse grained field is just given by the values of the original field on the coarse grained lattice, will lead to a local coarse grained action in one dimension.

\section{Coarse graining for systems with gauge symmetries}\label{Sec:CoarseGrainingWithGauge}

\subsection{Gauge degrees of freedom under coarse graining}\label{Sec:SubGaugeDOFUnderCoarse}

The formalism in section \ref{conv} can only be applied if the dynamics does not feature gauge symmetries, as otherwise the matrix $m_{AB}$ in the action (\ref{b1}) is not invertible. Of course one can perform a gauge fixing procedure, as is used for instance in \cite{Bietenholz:1995cy} for Yang Mills theory. We are here however interested in regaining gauge symmetries, hence we rather prefer to adopt a gauge invariant framework. Another advantage in doing so, is that topological field theories, i.e. those without propagating degrees of freedom, such as 2D electromagnetism and 3D gravity, will have form invariant actions under coarse graining.

Furthermore, a gauge fixing approach is not suitable for discrete gravity:
as mentioned in the introduction, discretizations of general relativity  usually break diffeomorphism symmetry \cite{bahrdittrich1}. Here we understand under a gauge symmetry the property, that  for given fixed boundary data the solutions of the theory are not unique. This characterization depends however on the kind of solution (specified by the boundary data) under consideration. Indeed in most discretizations of gravity, such as Regge gravity \cite{Regge}, flat space solutions are not unique.

The reason is the following: Regge calculus involves a discretization of space time by internally flat building blocks  -- in this case simplices. The metric information is encoded in the lengths of the edges of these building blocks. Curvature arises as  flat simplices might be glued together along a hinge -- an edge in three dimensions and a triangle in four dimensions -- such that the sum of the angles contributed by the glued simplices around this hinge differs from $2\pi$. This difference is the so--called deficit angle and measures the scalar curvature.

Flat space solutions can be constructed easily by triangulating flat space. To this end one just has to distribute a set of points and to connect all these points with (geodetic, that is straight) edges  so that one obtains a triangulation. The lengths of these edges are induced by the embedding flat geometry. Having one such flat triangulation with a determined set of edge lengths, one can obtain another flat triangulation (with a different set of edge lengths) by displacing any vertex of the first triangulation in the embedding flat geometry. This displacement will by definition not change the flatness of the geometry. Also it only changes the lengths of the edges adjacent to the vertex -- hence the change is only local and will in general not affect the boundary data. In this sense we obtain many gauge equivalent solutions -- for every internal vertex we obtain $d$ gauge parameters, where $d$ is the space--time dimension.

Basically we obtain gauge symmetries for the case of flat solutions as these can be mirrored exactly in the discretized theory. The same applies for homogeneously curved solutions (if a cosmological constant is present) if one uses homogeneously curved building blocks \cite{Bahr:2009qc, Bahr:2009qd}. Also here the gauge symmetries correspond to vertex displacements.

The gauge symmetries of the flat geometry survive if one considers linearized Regge calculus on such a flat background \cite{Rocek}. The gauge modes correspond to the infinitesimal change of the lengths variables induced by the displacement of vertices embedded in the flat background geometry. However this invariance is broken to higher order \cite{bahrdittrich1,Dittrich:2009fb}, that is the second order  gauge modes do appear in the higher than second order (potential) terms. This makes a perturbative expansion in general inconsistent: quantum mechanically one has to face the problem that modes appear in the higher potential terms for which however a propagator is missing. Even classically it turns out \cite{Dittrich:2009fb} that the higher order equations lead to non--linear consistency equations for the perturbative lower order (gauge) variables, including the one at zeroth order. That is the positions of the vertices in the flat background geometry, which is left to be arbitrary for the linearized theory, is fixed by the higher order perturbative equations.

One way to avoid these problems is to improve the action order by order. In this way one pushes the gauge breaking terms to higher and higher order. Although the linearized Regge action features (linearized) gauge symmetries, one even has to start with the improvement of the quadratic order of the action (defining the linearized theory). The reason is, that gauge breaking at third order is related to the non-invariance of the second order Hamilton--Jacobi functional of the theory under vertex displacements, see \cite{Dittrich:2009fb}. In other words the linearized theory although being invariant under infinitesimal vertex displacements is not invariant under finite vertex displacements and its predictions still depend on the underlying lattice.

What kind of diffeomorphism symmetry can one expect for the full non--perturbative perfect action? As this action should represent the pull--back of continuum physics to the lattice, we can describe the potential solutions of such an action. Assume that as in Regge calculus the basic variables are the lengths of the edges of some underlying triangulation. Then one way to obtain lattice representations of continuum solutions is to choose a triangulation of a given solution, i.e. to embed vertices in this solution and to connect these by geodetic edges. The geometry of the continuum solutions prescribes the length of these edges, determining a particular configuration of the lattice theory. Obviously there is a huge set of ambiguities in this procedure, namely the choice of how and where to embed the vertices into the continuum solutions. This is where a perfect lattice theory should lead to gauge equivalent solutions. That is, also non--perturbatively, one would expect vertex displacements as remnants of the continuum diffeomorphism symmetry. This can also be understood from the construction of the solutions described above: the choice of where to embed the vertices can be parametrized with the choice of coordinates. The change of coordinates under a transformation would thus induce a change of the embedded vertices and hence in general of the edge lengths describing the discrete solution.

Let us turn to the general problem of coarse graining theories with gauge symmetries. Conceptually this is not a problem at the classical level, as we can still apply the definition (\ref{defcg})
\ba\label{defcg2}
S'\left[\Phi\right]= \underset{ \phi,  \,B \phi=\Phi}{\text{extr}} \, S \q ,
\ea
i.e. to evaluate the action at an extremum under the conditions that the coarse grained fields $B\phi$ are equal to some prescribed values $\Phi$. In general this extremum will not be unique -- due to the gauge symmetries. But this does not render the coarse grained action (\ref{defcg2}) ill--defined, as by definition the values of the action at these gauge related extrema coincide.

Note also that gauge symmetries are preserved under coarse graining: If $\phi_s(\lambda)$ is a family of solutions related by gauge transformations labelled by $\lambda$, then -- as coarse grained solutions will be solutions of the coarse grained action -- $B\, \phi_s(\lambda)$ will be a family of solutions of the coarse grained action. What will in general happen is, that gauge degrees of freedom are absorbed by the coarse graining, i.e. that $B\, \phi_s(\lambda)$ is a much smaller set of solutions than $\phi_s(\lambda)$.

A useful criterion for the choice of the coarse graining map $B$ will be that it should preserve the form of the gauge symmetries for the coarse grained action, as will be discussed for the examples below. This will have the advantage that the (often geometric determined) interpretation of the gauge transformations will not change, nor does the form of the gauge invariant variables.

The coarse grained gauge modes can be easily described for free theories. Assume that the (symmetric) matrix $m_{AB}$ in the action\footnote{For this discussion we have absorbed the lattice labels $x$ or $p$  into the indices $A,B,\ldots$.}
\ba\label{g1}
S=\tfrac{1}{2} \sum_{A,B}  \phi_A \, m_{AB} \,  \phi_B    \q .
\ea
has null vectors $v_B$ such that $\sum_B  m_{AB} \,v_{B}=0$. We add the coarse graining conditions
\ba
\sum_{A'} \lambda_{A'} \left( \Phi_{A'}-\sum_B B_{A'B}\phi_B  \right)
\ea
 to the action (where the index $A'$ labels the coarse grained fields and will in general assume fewer values than the indices $A,B,\ldots$) and obtain the following equations of motion for the fields $\phi_A$ and the Lagrange multipliers $\lambda_A$
\ba\label{g2}
\sum_B m_{AB}\phi_B \;=\; \sum_{B'} \lambda_{B'} B_{B'A}  \q ,\q\q
\Phi_{A'} \;=\; \sum_B B_{A'B}\phi_B  \q .
\ea
As before we can write the coarse grained action as
\ba\label{g4}
S' &=&
 \frac{1}{2} \sum_{B'} \lambda_{B'} \Phi_{B'} \;=:\;  \frac{1}{2}\sum_{A',B'} \Phi_{A'}M_{A'B'} \Phi_{B'}
\ea
where $\lambda_{B'}$ is to be understood as a function of $\Phi_{A'}$ determined by the equations of motions (\ref{g2}). 
 Hence we define $M_{A',B'}$ to satisfy $\lambda_{A'}=\sum_{B'}M_{A'B'}\Phi_{B'}$.
Now  if $v_A$ is a null vector, we will have
 \ba\label{g6}
 0\;=\;\sum_{A,B} v_A m_{AB}\phi_B \;=\; \sum_{A'B} \lambda_{A'} B_{A'B}v_B \q .
 \ea
so that we obtain as a condition on $M$
 \ba\label{g7}
 \sum_{A',B}   (B_{A'B}v_B)\,M_{A'C'}=0 \q .
 \ea
 Therefore $V_{A'}=\sum_B B_{A'B}v_B$ is a null vector for the coarse grained action (\ref{g4}).

\subsection{Coarse graining of free theories with gauge symmetries}

Here we will derive a general formula for the coarse grained action in the case that gauge symmetries are present. We will directly work with the Fourier transformed fields, so that the action is
\ba\label{gg1}
S&=& \frac{1}{2 \,N^d}    \sum_{A,B} \sum_{p}    \phi_A(p)  \, \left( \Pi \cdot m(p) \cdot \Pi \right)_{AB} \phi_B(-p)
\ea
where we inserted projectors $\Pi_{CD}(p)$ onto the subspace orthogonal to the gauge modes, that is the null vectors of $m_{AB}(p)$.
 As before the coarse grained fields will be given as
\ba\label{gg2}
\Phi_A(P)&=&  \frac{1}{N^d} \sum_{B,p} B_{AB}(P,p) \, \phi_B(p)  \q .
\ea
Adding these conditions with Lagrange multipliers $\lambda(-P)$ to the action (\ref{gg1}) we will obtain the following equations of motion
\ba\label{gg3}
\sum_{B} \left(\Pi  \cdot m \cdot \Pi \right)_{AB}(p) \,\phi_B(-p) &=& \sum_{C}\sum_P  \lambda_C(-P)\, B_{CA}(P,p)   \\
\label{gg4}
\Phi_A(P)&=& \frac{1}{N^d}\sum_{B}\sum_p  B_{AB}(P,p)\,\phi_B(p)   \q .
\ea

By contracting the first equation with the projector $\Pi^\perp(p)$ onto the space of gauge modes $v^\alpha_a(p)$, labelled by an index $\alpha$, we learn that
\ba\label{gg5}
\sum_{A,C} \sum_P \,\lambda_C(-P) B_{CA}(P,p)\,\, \Pi^\perp_{DA}(p)\;=\;0 \q .
\ea
From the discussion in section \ref{Sec:SubGaugeDOFUnderCoarse} we know that $V^\alpha_A(P):=\sum_B\sum_p B_{AB}(P,p)v^\alpha(p)$ will be gauge modes of the coarse grained action. Let $\Pi^\perp_{AD}(P)$ be the projector onto the space spanned by these modes and $\Pi_{AD}(P)$ the projector orthogonal to $\Pi^\perp_{DE}(P)$. (In the examples below the projectors $\Pi^\perp(P),\Pi(P)$ will have the same form as $\Pi^\perp(p),\Pi(p)$ respectively, therefore we just use the same symbols here.) Hence equation (\ref{gg5}) entails
\ba\label{gg6}
\sum_C \lambda_C(-P) \,\,\Pi^\perp_{CA}(P)\;=\;0 \q .
\ea

Contracting the equation (\ref{gg4}) with the projector $\Pi(-P)$ we obtain
\ba\label{gg7}
N^d \,\, \sum_D \Pi_{AD}(-P)\cdot \Phi_D(P) &=& \sum_{B,C} \sum_P \,\Pi_{AB}(-P)\,B_{BC}(P,p) \phi_C(p) \nn\\
 &=& \sum_{B,C,D} \sum_P \,\,\Pi_{AB}(-P)\,B_{BC}(P,p) \Pi_{CD}(-p) \,\, \phi_D(p)
\ea
as any gauge modes in the field $\phi$ are projected away after coarse graining by $\Pi(P)$.

Let $m^{-g}_{AB}(p)$ be a generalized inverse to $m_{BC}(p)$. That is, $m^{-g}_{AB}(p)$ satisfies
\ba\label{gg8}
\sum_B m^{-g}_{AB}(p) \, m_{BC}(p)\;=\; \sum_B m_{AB}(p) \, m^{-g}_{BC}(p)\;=\; \Pi_{AC}(p)  \q .
\ea
The generalized inverse is not unique as one can add multiplies of the projector $\Pi^\perp$. These non--unique terms will however be projected out later on.
We can deduce from equation (\ref{gg3})
\ba\label{gg9}
\sum_B \Pi_{AB}(p) \,\phi_B(-p)&=& \sum_{BC} \sum_P m^{-g}_{AB}(p) \, B_{CB}(P,p) \,\lambda_C(-P)  \q,
\ea
which if used in (\ref{gg7}) yields
\ba\label{gg10}
N^d \,\, \sum_D \Pi_{AD}(-P)\, \Phi_D(P) \,=\!\!\!\! \sum_{B,C,D,E} \sum_{P,Q} \,\,\Pi_{AB}(-P)\,B_{BC}(P,p)   m^{-g}_{CD}(-p) \, B_{DE}(-Q,-p)  \,\lambda_E(Q). \nn\\
\ea
Because of equation (\ref{gg6}) we can replace $\lambda_E(Q)$ in the last equation (\ref{gg10}) by\\ $\sum_F \Pi_{EF}(-Q) \lambda_F(Q)$. We therefore have
\ba\label{gg11}
N^d \,\, \sum_D \Pi_{AD}(-P)\cdot \Phi_D(P) \,=\,  \sum_{D}\sum_Q M^{-g}_{AD}(P,Q) \,\lambda_D(Q)
\ea
with
\ba\label{gg12}
M^{-g}_{AB}(P,Q) =   \sum_{C,D,E,F} \sum_{p}    \,\,\Pi_{AC}(-P)\,B_{CD}(P,p)   m^{-g}_{DE}(-p) \, B_{EF}(-Q,-p)  \Pi_{FB}(-Q)  \q .
\ea
We now have to find a generalized inverse $M_{AB}(P,Q)$ to $M^{-g}_{BC}(Q,R)$ satisfying
\ba\label{gg13}
\sum_B \sum_Q  M_{AB}(P,Q)  M^{-g}_{BC}(Q,R) \;=\; \delta(P-R)\, \Pi_{AC}(-P)  \q .
\ea

As  in the section \ref{conv} the coarse grained action can be written as
\ba\label{gg14}
S'=\frac{1}{2} \sum_A\sum_P \Phi_A(P) \lambda_A(-P)=\frac{1}{2} \sum_{A,B}\sum_P \Phi_A(P)  \Pi_{AB}(P) \lambda_B(-P)
\ea
where $\lambda_A$ has to satisfy the equations of motion (\ref{gg3},\ref{gg4}). This solution is given by inverting (\ref{gg11}), hence the coarse grained action is given by
\ba\label{gg15}
S'=  \frac{N^d}{2} \sum_{A,B,C,D}\sum_{P,Q} \Phi_A(P)\,\, \Pi_{AB}(P)\, M_{BC}(-P,-Q)\, \Pi_{CD}(Q) \,\, \Phi_{D}(-Q)  \q .
\ea
The difference to the standard case without gauge symmetries (\ref{b16}) is, that we have to work with generalized inverses and that we have to insert the projectors $\Pi$ into the formula for the coarse grained action (\ref{gg12},\,\ref{gg15}). These projectors take care of the non--uniqueness of the generalized inverses, that is the coarse grained action $S'$ does not depend on the particular choice of representative for the generalized inverse.

\section{Example: Perfect action for electromagnetism}\label{Sec:Electromagnetism}

\subsection{Coarse graining from the lattice and the continuum }

\begin{figure}[hbt!]
    \begin{center}
	\psfrag{x}{$x$}	
	\psfrag{y}{$x+e_1$}	
	\psfrag{z}{$x+e_2$}	
	\psfrag{w}{}	
    \psfrag{a1x}{$a_1(x)$}
    \psfrag{a2x}{$a_2(x)$}
    \psfrag{a2y}{$a_2(x+e_1)$}
    \psfrag{a2z}{$\!\!\!\!\!\!a_1(x+e_2)$}
\includegraphics[scale=0.8]{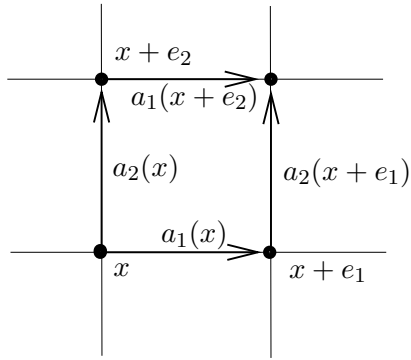}
    \caption{The variables used in discrete electromagnetism (here for $d=2$).}\label{Fig:2dEMVar}
    \end{center}
\end{figure}

Here we will first discuss electromagnetism (or Abelian Yang Mills fields), as this is a much simpler example for a lattice theory with gauge symmetries than lattice  gravity. The basic fields will be connection variables $a_b$ associated to the edges of the lattice. Here $a_b(x)$ is the variable associated to the (positively oriented) edge starting at the site $x$ in the direction $b$ (see figure \ref{Fig:2dEMVar}).

A discretization for the action is given by the square of plaquette variables $f_{ab}$
\ba\label{em1}
S=\frac{a^{d-2}}{2} \sum_x \sum_{b<c} f_{bc} (x) f_{bc}(x)
\ea
 where
 \ba\label{em2}
 f_{bc}(x)=a_b(x)+ a_c(x+e_b) -a_b(x+e_c)-a_c(x)  \q
 \ea
and $a$ is as before the lattice constant.
The Fourier transformed plaquette variable is given by
\ba\label{em3}
f_{bc}(p) &=& a_b(p) + e^{2\pi i \frac{p_b}{N} } a_c(p)- e^{2\pi i \frac{p_c}{N} } a_b(p)-a_c(p) \nn\\
&=& k_c a_b - k_b a_c
\ea
and the action as
\ba\label{em4}
S&=&\frac{a^{d-2}}{2 N^d}\sum_p\sum_{b<c} f_{bc}(p)  \, f_{bc}(-p) 
\;=\; \frac{a^{d-2}}{2 N^d}\sum_p   \sum_{b,c}\,\, a_b(p)\,\Delta' \left(  \delta_{bc} - \frac{\bar{k}_b k_c}{\Delta'}\right) a_c(-p)
\ea
where $\Delta'=\sum_b k_b \bar{k}_b= a^2 \Delta$. Here we can introduce the discretized projectors onto the transversal $\Pi^t$ and longitudinal $\Pi^l$ modes
\ba\label{em5}
\Pi_{bc}^{t}=\delta_{bc}-\frac{\bar{k}_b k_c}{\Delta'}\q, \q\q \Pi^{l}_{bc}=\frac{\bar{k}_b k_c}{\Delta'}
\ea
satisfying
\ba\label{em6}
\sum_{c} \Pi_{bc}^\alpha \Pi_{cd}^\beta=\delta^{\alpha\beta}  \Pi_{bd}  \q , \q\q \Pi_{bc}^{t}+\Pi^{l}_{bc}=\delta_{bc}
\ea
for $\alpha,\beta=t,l$. (Note that here the projectors are meant to act on $a_c(-p)$ on the right and $a_b(p)$ on the left.) The action is therefore a sum over only the transversal modes -- the longitudinal modes $a_b \sim k_b$ do not appear and are hence gauge modes. This corresponds to the gauge symmetry
\ba\label{em7}
a_b (x) \mapsto a_b(x)+g(x+e_b)-g(x)
\ea
with a gauge parameter $g(x)$ at each lattice site $x$.

\begin{figure}[hbt!]
    \begin{center}
	\psfrag{x}{$X$}	
	\psfrag{a1}{$a_1(x)$}	
	\psfrag{a2}{$a_2(x)$}	
    \psfrag{b1}{$A_1(X)$}
    \psfrag{b2}{$A_2(X)$}
\includegraphics[scale=0.4]{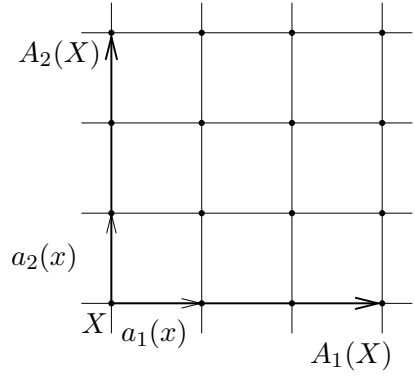}
    \caption{The variables in electromagnetism are coarse grained along the lines of the lattice, since they are naturally one-forms. (here with $D=2$ and $L=3$)}\label{Fig:2dEMVarCoarse}
    \end{center}
\end{figure}

Let us turn to the coarse graining of the fields. The connection is a one--form - hence naturally discretized as variables associated to edges (see figure \ref{Fig:2dEMVarCoarse}).

Coarse graining would mean to integrate the connection over all the (smaller) edges that built up the new (longer) edge. Hence we define
\ba\label{em8}
A_c(X)= b \sum_z a_c(LX+ z e_c) &=:&   \sum_{b,x}   B_{cb}(X,x)\, a_b(x)
\ea
where $z=0,\ldots,L-1$ and $b$ is a rescaling factor.  The Fourier transformed coarse graining matrix is then
\ba\label{em9}
B_{cd}(P,p)=    b\, N'^d  \,\,\delta_{cd} \,\, \delta^{(N')}(P-p)\,\,    \frac{K_c}{k_c} \q .
\ea
As can be easily seen the coarse graining operation has the remarkable property that it transforms the longitudinal modes $a_c \sim k_d$ , which are the gauge modes of the action,  to longitudinal modes on the coarse grained lattice $A_d \sim K_d$.
Hence these modes will be also gauge modes of the coarse grained action.
The coarse grained variables keep their geometric interpretation: for instance the coarse grained plaquette variables $F_{bc}=K_cA_b-K_bA_c$ will be invariant under the gauge transformations of the coarse grained action.

As the longitudinal modes are preserved by the coarse graining we will have
\ba\label{em10}
\sum_{d,e}\sum_p   B_{cd}(P,p)\,\, \Pi^{l}_{de}(-p) \,\, B_{fe}(-Q,-p) &=& b^2 N'^{2d} \, \delta(P-Q) K_c \bar{K}_f
 \sum_r \frac{1}{\Delta'(p)}_{|p=P+N'r} \nn\\
 &\sim&  \Pi^{l}_{cf}(-P) \q\q\q
\ea
where in the sum $r$ takes values $r=(0,\ldots,0),\ldots,(L-1,\ldots,L-1)$. This will allow us to add an arbitrary multiple of the longitudinal projector to the generalized inverse $m^{-g}$ in formula (\ref{gg8}) for the coarse graining, as this added part will be projected out again by the transversal projectors. Hence, we use  for the generalized inverse
\ba\label{em11}
(m^{-g})_{cd} \,= \,\Delta'^{-1} P^{t}_{cd} \,\simeq \,\Delta'^{-1} \delta_{cd}  \q .
\ea
where the last equality holds modulo terms proportional to the longitudinal projector.
This gives for the matrix $M^{-g}$ appearing in the  coarse grained action
\ba\label{em12}
(M^{-g})_{ch}(P,Q)&=& \sum_{d,e,f,g}\,\,
\Pi^{t}_{cd}(-P)   \left(\sum_p B_{de}(P,p) \,\, (m^{-g})_{ef}(-p)\,\, B_{gf}(-Q,-p)\right)  \Pi^{t}_{gh}(-Q)
\nn\\
&=& b^2 N'^{2d}\,\,  \delta^{(N')}\!(P-Q)\,\,\,\,
\Pi^{t}_{cd}(-P) \,\,  \,\,s_d\,  K_{d} \bar{K}_d \,\,\delta_{dg}   \, \,\,\,  \Pi^{t}_{gh}(-Q)
\ea
 where
 \ba\label{em13}
  s_d&=& \sum_r  \left(\frac{1}{\Delta'(p)} \frac{1}{k_d \bar{k}_d}\right) _{\big{|}p=P+N'r}   \q .
\ea

The generalized inverse can be found by adding a longitudinal part of the form
\be\label{em14}
\lambda \,\,   b^2 N'^{2d}\,\,  \delta^{(N')}\!(P-Q)\,\,\,\,  \Pi^{l}_{cd}(-P) s_d  K_{d} \bar{K}_d \,\,\delta_{dg}   \,   \Pi^{l}_{gh}(-Q)
\ee
 and to invert the sum of the terms. Projecting from both sides with $\Pi^{t}(-P)$ gives a generalized inverse which is independent of $\lambda$, satisfying
  \be
 \sum_h\sum_Q M^{-g}_{ch}(P,Q) M_{hf}(Q,R)=\delta^{(N')}(P-R)\,\Pi^t_{cf}(-P)  \q .
\ee
 In this way we obtain
 \ba\label{em15}
 M_{cd}(P,Q)=   \frac{1}{b^2 N'^{2d}} \,\, \delta^{(N')}(P-Q) \,\, \frac{1}{\bar{K}_c {K}_d} \frac{1}{ t} \left(
 \delta_{cd}\,t_c - (1-\delta_{cd})  t_{cd} \right)
 \ea
 where
 \ba\label{em16}
 t_{cd}=\prod_{e \neq c,d} s_e \q,\q\q
 t_c=\sum_{e\neq c} t_{ce} \q,\q\q
 t= \sum_e \, \prod_{f\neq e} s_f   \q .
 \ea

Finally, the coarse grained action is given by
 \ba\label{em17}
 S' &=& \frac{a^{d-2}N^d}{2} \sum_{P,Q} A_c(P) M_{cd}(-P,-Q) A_d(-Q)  \nn\\
 &=&
 \frac{1}{2}\, \frac{L^d\,a^{d-2}}{N'^d b^2} \sum_{c,d} \sum_P A_c(P) \,\, \frac{ \left(\delta_{cd}t_c-(1-\delta_{cd})t_{cd}\right)}{t\,\,K_c \bar{K}_d} \,\,A_d(-P) \q .
 \ea
 It is straightforward to see  --
 using $\sum_c \left(t_c\,\delta_{cd} -  t_{cd} (1-\delta_{cd})\right) =0$ -- that in the coarse grained action
 the longitudinal modes $A_b(P)\sim K_b$ are indeed gauge modes.

~\\

To `block from the continuum' we proceed with the same conventions as for the scalar field in section \ref{blockcont}. Accordingly, we start from the continuum action
\ba
S_c&=& \frac{1}{2}\sum_{b<c}\int_{[0,T)^d} \bd^d\chi\, \left(\partial_b a_c(\chi) -\partial_c a_b(\chi)\right)^2
\nn\\
 &=&
\frac{1}{2 T^{d+2}} \sum_{b,c}\sum_{\kappa \in \mathbb{Z}^d} a_b(\kappa)\left( \sum_d (2\pi \kappa_d)(2\pi\kappa_d)\delta_{bc}-(2\pi \kappa_b) (2\pi \kappa_c) \right)a_c(-\kappa) \q .
\ea
The coarse grained connection variables $A_b$ are obtained by integrating the connection $a_b$ over the edges of the lattice:
\ba
A_b(X)&=&b_c \int_{[0,a')}\bd^d \epsilon_b \,\,a_b(a'X+\epsilon_b)\q , \q \q\q\q\q\q\text{so that} \nn\\
A_b(P)&=&\frac{b_c N'^d}{T^{d-1}} \sum_{\kappa \in \mathbb{Z}^d} \,\delta^{(N')}(P-\kappa)\, a_b(\kappa) \left[  \frac{K_b}{\frac{2\pi}{i}\kappa_b}\right] \, .
\ea
This gives for the coarse grained action
\ba
S'&=& \frac{1}{2}\,\frac{T^{d-4}}{N'^{2d} b_c^2} \sum_{c,d,P} A_c(P)  \,\, \frac{ \left(\delta_{cd}t_c-(1-\delta_{cd})t_{cd}\right)}{t\,\,K_c \bar{K}_d} \,\,A_d(-P) \; ,
\ea
where $t,t_c,t_{cd}$ are defined as before, equation (\ref{em16}), just that $s_d$ is now given by
\ba
s_d&=& \sum_{r\in \mathbb{Z}^d}  \frac{1}{(2\pi)^4} \left( \frac{1}{\sum_b \kappa_b\kappa_b}\,\frac{1}{\kappa_d \kappa_d}\right)_{\big{|} \kappa=P+N'r}  \q .
\ea

\subsection{Electromagnetism in two dimensions}\label{Sec:Sub2DElectromagnetism}

Here we will consider the two--dimensional case and show that the action is form invariant under coarse graining.
For the quantities appearing in (\ref{em16}) we have
\ba\label{em18}
t_{12}=1 \q , \q\q t_c=1\q , \q\q t=s_1+s_2   \q .
\ea
For the last quantity we obtain with the definition (\ref{em13})
\ba\label{em19}
s_1+s_2&=& \sum_r  \frac{1}{k_1 \bar{k}_1 + k_2 \bar{k}_2}\big( \frac{1}{ k_1 \bar{k}_1} +\frac{1}{k_2\bar{k}_2} \big) \nn\\
&=& \sum_r  \frac{1}{k_1 \bar{k}_1 \, k_2 \bar{k}_2} \nn\\
&=& L^4 \frac{1}{K_1 \bar{K}_1 \, K_2 \bar{K}_2}
\ea
where the last identity is proved in  appendix \ref{somesums}. Note that the summation over $r$ for the expression $s_1+s_2$ just replaces the fine grained wave vectors $k$ with the coarse grained ones $K$ (and introduces a factor of $L^4$). This is equivalent to considering a coarse graining step with $L=1$ where $k=K$, so we will indeed find, that the action just undergoes a rescaling if coarse grained.

The coarse grained action is given by
\ba\label{emEnd}
S' &=& \frac{1}{2N'^2}\frac{1}{b^2 L^2}  \sum_{c,d} \sum_P A_c(P) \,\, \frac{K_1 \bar{K}_1 K_2 \bar{K}_2}{K_c \bar{K}_d} \left(2 \delta_{cd}-1\right)\,\,A_d(-P) \nn\\
&=&
 \frac{1}{ 2 N'^2} \frac{1}{b^2 L^2}  \sum_{c,d} \sum_P A_c(P) \,\,
\Delta'_K\, \left( \delta_{cd}-  \frac{\bar{K}_c K_d}{\Delta'_K}\right)
  \,\,A_d(-P)
\ea
where we abbreviated $\Delta'_K=\sum_d K_d \bar{K}_d$.

Indeed the action (\ref{emEnd}) is a multiple of the action we started with (\ref{em4}).   The form invariance of the action can be easily understood if one works with the plaquette variables $f_{ab}$ as in this case the matrix $m$ appearing in the action (\ref{em1}) is just the identity. This does not apply to higher dimensions, as  the plaquette variables are not (locally) independent anymore, due to the Bianchi identities \cite{Batrouni:1984rb}. Nevertheless this shows, that looking for variables which are particularly convenient for coarse graining might very much simplify the calculations \cite{toappear}.

\section{Example: Linearized gravity}\label{Sec:LinearizedGravity}

\subsection{The  coarse graining}

Next we will consider linearized gravity discretized on a lattice. In this section we will supply all the necessary ingredients to perform the coarse graining, that is a discrete action, including a discretization of spin--0, spin--1 and spin--2 projectors for the metric variables, and a geometrically derived coarse graining map. We will then consider 3D linearized gravity and show that the discrete action is invariant under coarse graining. This has to be expected as 3D gravity is a topological theory, i.e. there are no propagating degrees of freedom.

\begin{figure}[hbt!]
    \begin{center}
	\psfrag{a}{$e_a$}	
	\psfrag{b}{$e_b$}	
	\psfrag{c}{$e_c$}	
    \psfrag{bc}{$e_{bc}$}
    \psfrag{abc}{$e_{abc}$}
\includegraphics[scale=0.75]{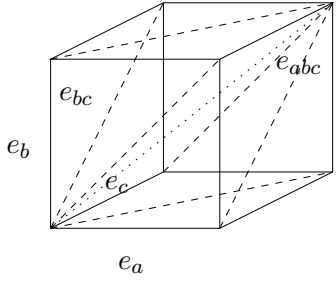}
    \caption{For gravity, the lattice (solid lines) needs to be enhanced by introducing (e.g. for $D=3$) face-diagonals (dashed lines) and body-diagonals (dotted lines) in order to capture all triangulation degrees of freedom of Regge calculus.}\label{Fig:Hypercube}
    \end{center}
\end{figure}

The choice of discretization for (linearized) gravity is not as straightforward as in the case of electromagnetism. One popular example is provided by Regge gravity \cite{Regge}. Regge gravity relies on a triangulation with basic variables given by the lengths of the edges (see figure \ref{Fig:Hypercube}). See \cite{barrett} for a discussion under which conditions linearized Regge solutions do converge to linearized (continuum) gravity solutions. Hence we can hope to recover continuum physics and therefore symmetries in the coarse grained action by taking the fine lattice, we coarse grain from, to the continuum limit.

Choosing a regular (hyper--) cubical lattice \cite{Rocek} linearized Regge gravity can be mapped to linearized gravity with fundamental variables given by symmetric tensors $h_{ab}$ associated to the vertices of the lattice. These variables represent the metric perturbations from flat space. The map is reviewed in appendix \ref{appmetric} since it will be used to derive the coarse graining map for the variables $h_{ab}$.

~\\

To define the discrete action for linearized gravity, let us first consider the continuum Lagrangian density in $d=3,4$ dimensions
\ba\label{grav1}
L_{cont}= \frac{1}{4}  \sum_{a,b,c,d}  \, h_{ab}\,  \left(\Pi^2-(d-2)\Pi^0 \right)_{abcd} \,\Delta \, \, h_{cd}   \q
\ea
where $\Delta=- \sum_a \partial_a \partial_a \equiv  \sum_a k_a k_a$.

Here we introduced the spin projectors
\ba\label{grav2}
\Pi^0_{abcd}  &=&  \frac{1}{d-1} \Pi^{t}_{ab} \Pi^{t}_{cd}  \nn\\
\Pi^1_{abcd} &=& \frac{1}{2}(\delta_{ac}\delta_{bd}+\delta_{ad}\delta_{bc}) \,-\, \frac{1}{2}(\Pi^{t}_{ac}\Pi^{t}_{bd}+\Pi^t_{ad}\Pi^t_{bc}) \nn\\
\Pi^2_{abcd}&=& \frac{1}{2}(\Pi^{t}_{ac}\Pi^{t}_{bd}+\Pi^{t}_{ad}\Pi^{t}_{bc}) - \frac{1}{d-1} \Pi^{t}_{ab} \Pi^{t}_{cd}
\ea
where
\ba
\Pi^{t}_{ab} =\delta_{ab}-\frac{k_a k_a}{\Delta}  \equiv  \delta_{ab}+ \frac{\partial_a \partial_b}{\Delta}
\ea
is the projector onto the transversal modes. As can be easily seen $\Pi^0,\Pi^1$ and $\Pi^2$ sum to the identity map (on the space of symmetric rank two tensors)
\ba
(\Pi^0+\Pi^1+\Pi^2)_{abcd}= \frac{1}{2}(\delta_{ac}\delta_{bd}+\delta_{ad}\delta_{bc})  \q .
\ea
and are orthogonal to each other.

In the gravity Lagrangian (\ref{grav1}) the projector $\Pi^1$ does not appear, hence the modes it projects on are gauge modes. More precisely the longitudinal modes
\ba
v_{ab}^c=\delta^c_a k_b+ \delta^c_b k_a
\ea
are annihilated by $M_{abef}:=\frac{1}{2}( \Pi^2-(d-2)\Pi^0 )_{abef} $ for $c=1,\ldots,d$.

~\\

We will now discuss the discrete action. To this end we will define discretize projectors and replace the continuum Laplacian by the lattice Laplacian (\ref{b19}). This is most easily done in the Fourier transformed picture. Here the difference to the continuum is, that we have the choice between $k_a=1-e^{\frac{2\pi i}{N}p_a}$ and $\bar{k}_a=1-e^{\frac{-2\pi i}{N}p_a}$, that is forward and backward lattice derivative. But we also have to satisfy a discretization condition, which is that the projectors should be Hermitian, i.e. $\Pi^i_{abcd}=\bar{\Pi}^i_{cdab}$.  Hence we define
\ba\label{disreteprojectors}
\Pi^0_{abcd} &=&\frac{1}{d-1}
\left(\delta_{ab}+\frac{\bar{k}_a \bar{k}_b}{\Delta'} (1-\delta_{ab} {k}_b) \right)  \,\,
 \left(\delta_{cd}+\frac{k_c k_d}{\Delta'} (1-\delta_{cd}\bar{k}_d) \right)  \,\,
      \nn\\
 \Pi^1_{abcd}  &=&   \frac{1}{2}(\delta_{ac}\delta_{bd}+\delta_{ad}\delta_{bc})-\nn\\
 &&\frac{1}{2} (1-\delta_{ab}{k}_b)(1-\delta_{cd}\bar k_d)
 \left(
 (\delta_{ac}-\frac{\bar k_a {k}_c}{\Delta'}) \,(\delta_{bd}-\frac{\bar k_b {k}_d}{\Delta'}) +
 (\delta_{ad}-\frac{\bar k_a {k}_d}{\Delta'}) \,(\delta_{bc}-\frac{\bar k_b {k}_c}{\Delta'})
 \right)
 \nn\\
 \Pi^2_{abcd}&=&
 \frac{1}{2} (1-\delta_{ab}{k}_b)(1-\delta_{cd}\bar k_d)
 \left(
 (\delta_{ac}-\frac{\bar k_a {k}_c}{\Delta'}) \,(\delta_{bd}-\frac{\bar k_b {k}_d}{\Delta'}) +
 (\delta_{ad}-\frac{\bar k_a {k}_d}{\Delta'}) \,(\delta_{bc}-\frac{\bar k_b {k}_c}{\Delta'})
 \right) -\nn\\
 && \frac{1}{d-1}  \left(\delta_{ab}+\frac{\bar{k}_a \bar{k}_b}{\Delta'} (1-\delta_{ab} {k}_b) \right)  \,\,
 \left(\delta_{cd}+\frac{k_c k_d}{\Delta'} (1-\delta_{cd}\bar{k}_d) \right) \q
 \ea
 where $\Delta'=\sum_a k_a \bar{k}_a$. Note that
\be
k_a(1-\bar{k}_a)= -\bar{k}_a \q, \q\q \bar{k}_a(1-k_a)=-k_a \q, \q\q (1-k_a)(1-\bar{k}_a)=1
\ee
so the additional factors $(1-\delta_{ab}{k}_b),(1-\delta_{ab}\bar{k}_b)$  just change forward derivative into backward derivatives and vice versa.

The discrete action so obtained
\ba\label{gravaction}
S=\frac{1}{4 N^d}  \sum_{a,b,c,d} \sum_p \, h_{ab}(p)\,  \left(\Pi^2-(d-2)\Pi^0 \right)_{abcd}\, \Delta'  \,\, h_{cd}(-p)
\ea
reproduces the linearized Regge action derived in \cite{Rocek} on a regular (hyper--) cubical lattice. The gauge modes for this action
\ba
v^c_{ab}(p)=\delta^c_a k_b (1 +(\delta^c_b-1)k_a) + \delta^c_b k_a (1 +(\delta^c_a-1)k_b)
\ea
correspond to the change in the metric perturbation variables (via the change in the lengths of the edges of the triangulation) if a vertex is infinitesimally displaced in the triangulation \cite{Rocek}.

We now turn to the coarse graining map for the fields $h_{ab}$. Here it is important to use the geometric nature of the variables - namely that these encode edge lengths - to define a coarse graining map. As we will see, this ensures that coarse graining preserves the gauge modes. In appendix \ref{appmetric}  we derive the coarse graining for the $h_{ab}$ induced by the natural coarse graining for the edge lengths - namely that a coarse grained edge length is just the sum of the length of the edges contained in the coarse grained edge.  The resulting map is given by
\ba\label{gravv1}
B_{abcd}(P,p)\;=\;\frac{N^d}{L^{d-1} }\delta^{(N')} (P-p)\, \Bigg(\!\!\!\! \!\!\!\!&&
\delta_{ab}\delta_{cd} \; \; \delta_{ac} \frac{K_a}{k_a}
+
\nn\\
&&(1-\delta_{ab})\delta_{cd} \;\frac{1}{2}\left( \delta_{ac} \left(\frac{K_{ab}}{k_{ab}} -\frac{K_a}{k_a}\right) + \delta_{bc} \left(\frac{K_{ab}}{k_{ab}} -\frac{K_b}{k_b} \right)\right) + \nn\\
&&
 (1-\delta_{ab})(1-\delta_{cd}) \;\frac{1}{2}(\delta_{ac}\delta_{bd}+\delta_{ad}\delta_{bc}) \frac{K_{ab}}{k_{ab}}
\Bigg)  \q ,
\ea
where we abbreviated $k_{ab}=1-e^{\frac{2\pi i}{N}p_a} e^{\frac{2\pi i}{N}p_b} =k_a+k_b-k_a k_b$ and similarly for $K_{ab}$. Note that this coarse graining is much more complicated than the one for electromagnetism (\ref{em9}) as we have now to deal with a non--diagonal matrix:  the second line in (\ref{gravv1}) displays non--vanishing entries between non--diagonal metric elements $H_{ab}$ and diagonal metric elements $h_{cc}$ .

The coarse graining matrix preserves however the gauge modes, that is
\ba
\sum_{cd} \sum_p B_{abef} \,v^c_{ef}(p) \;\sim  \;v^c_{ab}(P)  \q .
\ea
This justifies the choice of the coarse graining (\ref{gravv1}).

We now have all the necessary ingredients to apply the general formalism in section \ref{Sec:CoarseGrainingWithGauge}. The calculations are however considerably more involved than for electromagnetism. We leave the 4D case for future work and consider in the following subsection the 3D case, where the action will be invariant under coarse graining. This will show that the method, the discrete action as well as the coarse graining map proposed here lead to sensible results.

\subsection{Three--dimensional linearized gravity}\label{Sec:3DLinearizedGravity}

To find the coarse grained action we have to consider the following matrix according to the general formalism developed in section \ref{Sec:CoarseGrainingWithGauge}
 \ba\label{novi1}
M^{-g}_{abcd}= \sum_{e,e',f,f',g,g',h,h'} \sum_{p}    \,\,(\text{Id}-\bar \Pi^1)_{abef} \,\,B_{efgh}  \,\,  \bar{m}^{-g}_{ghg'h'} \, \,\bar{B}_{e'f'g'h'}\,\,(\text{Id}-\bar{\Pi}^1)_{e'f'cd}  \q .
\ea
Here, to keep the formulas readable, we surpressed the dependence on the momentum labels $p,P,Q$ and used a bar to indicate an object depending on $-p,-P$ instead of $p,P$ respectively. That is $\bar{\Pi}^1=\Pi^1(-P)$ and so on. The generalized inverse $\bar m^{-g}$ can be easily found using the representation of the matrix $m$ in (\ref{gravaction}) with projectors. Hence
\ba
\bar{m}^{-g}_{abcd} &=& 2 \,\,  \frac{1}{\Delta'}\,\left(\bar\Pi^2- \bar\Pi^0 \right)_{abcd}  \q .
\ea

It turns out, that the matrix elements $M^{-g}_{abcd}$ can be computed and have the following general form
 \ba\label{novi3}
M_{abcd}^{-g}\;=\;\sum_p\frac{A(k)}{\Delta_k'} \frac{\left[C_{abcd}^{12}(K) k_1k_2+ C_{abcd}^{13}(K)
k_1k_3 +C_{abcd}^{23}(K) k_2k_3+C_{abcd}^{123}(K)
k_1k_2k_3\right]}{\left[\Delta_K'(1-K_1)(1-K_2)(1-K_3)\right]^4}\,,\q\q\q
\ea where $\Delta_k'=\sum_a k_a \bar{k}_a$ and $\Delta_K'=\sum_{a}K_a\bar K_a$.
The prefactor $A(k)$ is given by
\ba
A(k)=2 \, \frac{N^6}{L^4} \delta^{(N')}(P-p)\,\delta^{(N')}(Q-p)\,\,\, \frac{(1-k_1)(1-k_2)(1-k_3)\Delta'_k}{k_1k_2k_3k_{12}k_{13}k_{23}},
\ea
with $k_{ab}=k_a+k_b-k_ak_b$. Note that there is a Laplacian in $A(k)$ that cancels the one in the denominator of (\ref{novi3}). This will eventually make the sum over $p$ in (\ref{novi3}) computable. The
coefficients $C_{abcd}$'s are polynomials of $K_1,K_2,K_3$, for example
\ba
C_{1111}^{12}&=& - C_{1111}(K) \,\big[ K_2 K_3 (1-K_1) (1-K_2) (1-K_3) \left( K_1\bar K_1 - K_2 \bar K_2  -K_3\bar K_3\right)\,+  \nn\\ &&\q\q\q\q\q \,\, K_1 K_3^2 (1-K_2) \big( 1- (1-K_1)((1-K_2)(1-K_3)\big)\big]   \nn\\
C_{1111}^{13}&=& - C_{1111}(K) \,\big[ K_2 K_3 (1-K_1) (1-K_2) (1-K_3) \left( K_1\bar K_1 - K_2 \bar K_2  -K_3\bar K_3\right)\,+  \nn\\ &&\q\q\q\q\q \,\, K_1 K_2^2 (1-K_3) \big( 1- (1-K_1)((1-K_2)(1-K_3)\big)\big]   \nn\\
C_{1111}^{23}&=&C_{1111}(K)\,K_1K_{23} (1-K_2)(1-K_3) \big[ K_2\bar K_2 + K_3 \bar K_3 -  K_1  K_2 -  K_1  K_3\big]
\nn\\
C_{1111}^{123}&=& C_{1111}(K) K_2K_3(1-K_1)(1-K_2)(1-K_3)\big[2K_1\bar K_1-K_2\bar K_2-K_3\bar K_3\big]
\ea
where the common factor $C_{1111}$ is given by
\ba
C_{1111}(K)&=& 2 K_1^2 K_2K_3 (1-K_1)^2(1-K_2)^2(1-K_3)^2 \big[ K_2\bar K_2 + K_3\bar K_3\big] \q .
\ea
Using an algebra manipulation program, the expressions for all the coefficients
have been obtained, but they are too long to be listed here.


The sum over the labels $p$ in (\ref{novi3}) involves only the $k_a=1-e^{2\pi i \frac{p_a}{N}}$ and not the $K_a=1-e^{2\pi i \frac{P_a}{N'}}$. We therefore have only to consider two types of sums, for instance
\ba
&&\sum_p \delta^{(N')}(P-p)\,\delta^{(N')}(Q-p)\,\,\, \frac{(1-k_1)(1-k_2)(1-k_3)}{k_1k_2k_3k_{12}k_{13}k_{23}}\,k_1k_2k_3  \nn\\ &=&\delta^{(N')}(P-Q)\sum_{r_1,r_2,r_3}
{ \frac{(1-k_1)(1-k_2)(1-k_3)}{k_{12}k_{13}k_{23}} \; }_{\big{|}k_a=1-e^{2\pi i \frac{P_a}{N}+2\pi i\frac{r_a}{L}}} \q\q .
\ea
The derivation of these sums can be found in appendix \ref{somesums}, the results are given by
\ba\label{sumsnov} \sum^{L-1}_{r_1,r_2,r_3=0}\!\!\frac{(1-k_1)(1-k_2)(1-k_3)}{
k_{12}k_{13}k_{23}}\!&=&\!\frac{(1-K_1)(1-K_2)(1-K_3)}{
K_{12}K_{13}K_{23}}\times\begin{cases}
  0   \q\q \;
 &  \mbox{if} \; L\; \mbox{is even} \\
 ~\\
  L^3         &   \mbox{if}\; L\; \mbox{is odd.}  \q\q\q
\end{cases}\nn\\
\sum^{L-1}_{r_1,r_2,r_3=0}\!\!\frac{(1-k_1)(1-k_2)(1-k_3)}{k_a \,\,
k_{12}k_{13}k_{23}}\!&=&\!\frac{(1-K_1)(1-K_2)(1-K_3)}{
K_{12}K_{13}K_{23}}\times\begin{cases}
  \frac{L^4}{2} \frac{2-K_a}{K_a}     \q\q \;
 &  \mbox{if} \; L\; \mbox{is even} \\
 ~\\
  \frac{L^4}{2}    \frac{2-K_a}{K_a}
  +  \frac{L^3}{2}         &   \mbox{if}\; L\; \mbox{is odd.} \q\q\end{cases}  \!\!\!\!\!\!\!\!\!\!\!\!\!\! \nn\\
\ea
Here we have two scalings, some terms scale with $\sim L^3$ (only for odd $L$) the others with $\sim L^4$. The terms with $L^3$--scaling do however vanish if we use the expressions for the coefficients $C^A_{abcd},\,A=12,13,23,123$: Namely to evaluate the $L^3$ terms for odd $L$ we have to consider
\be\label{novia}
C_{abcd}^{12}+C_{abcd}^{13}+C_{abcd}^{23}+2C_{abcd}^{123} \;=\;0,\ee
which vanishes for all index combinations $a,b,c,d$. We are thus left with the $L^4$--scaling terms in the sums (\ref{sumsnov}) which do agree for odd and even $L$. These terms sum up to
\ba\label{novib}
\frac{C_{abcd}^{12}\,\frac{2-K_3}{2K_3}+C_{abcd}^{13}\,\frac{2-K_2}{2K_2}+C_{abcd}^{23}\,\frac{2-K_1}{2K_1}}{   \,(\Delta_K')^4\,\,K_{12}K_{13}K_{23}(1-K_1)^3(1-K_2)^3(1-K_3)^3}=\frac{
\left(\bar\Pi^2(K)-\bar\Pi^0(K)\right)_{abcd}}{(\Delta_K')}\q ,
\ea
i.e. we obtain back a multiple of the original matrix $m_{abcd}$ we started with.
 In addition, by combining the two equations (\ref{novia},\ref{novib}) together,
we obtain the equation which guarantees the consistency
condition for $L=1$ so that $K_a=k_a$ for $a=1,2,3$. This case -- since no proper coarse graining is taken place -- should result  in the original matrix we started with. Indeed
 \be
 \frac{C_{abcd}^{12}\,\frac{1}{K_3}+C_{abcd}^{13}\,\frac{1}{K_2}+C_{abcd}^{23}\,\frac{1}{K_1} + C_{abcd}^{123}}{   \,(\Delta_K')^4\,\,K_{12}K_{13}K_{23}(1-K_1)^3(1-K_2)^3(1-K_3)^3}=\frac{
\left(\bar\Pi^2(K)-\bar\Pi^0(K)\right)_{abcd}}{(\Delta_K')} \q .
\ee

~\\

Collecting all prefactors we obtain for the coarse grained action
\ba
S'&=&\frac{1}{4N^3} \sum_{a,b,c,d}\sum_P H_{ab}(P)\,\left(\Pi^2-\Pi^0\right)_{abcd}\,\Delta'_K \, \, H_{cd}(-P)
\ea
showing that the 3D discrete linearized gravity action is indeed invariant under coarse graining.

Note that by comparison it is quite straightforward to derive the topological character of 3d (non--perturbative) Regge gravity, see also \cite{Bahr:2009qc}. The action in this case is
\ba\label{regge1}
S=\sum_e l_e \epsilon_e(l)
\ea
where the sum is over all edges in the triangulation, $l_e$ denotes the length of an edge $e$, and $\epsilon_e(l)$ is the so--called deficit angle, a measure for the curvature, associated to the edge $e$.  The fundamental variables are the edge lengths $l_e$.

We choose a coarse grained triangulation, such that the edges $E$, triangles and tetrahedra of the coarse grained triangulation are made up of the edges, triangles and tetrahedra, respectively, of the original triangulation. Then an obvious choice for the coarse graining map is to require that the length of a new edge $L_E$ is equal to the sum of the lengths of the edges $e$ contained in $E$. We add these conditions to the Regge action (\ref{regge1}) and therefore have to vary
\ba\label{regge2}
S_\lambda= \sum_e l_e \epsilon_e(l) +\sum_E \lambda_E \left(L_E-\sum_{e\subset E} l_e\right)  \q .
\ea
We obtain the equations of motion\footnote{Here one has to use the Schl\"afli identity \cite{Regge} to find that the terms with derivatives of the deficit angles cancel each other. This is equivalent to finding in the continuum, that the equations of motion only involve second order derivatives of the metric.}
\ba\label{regge3}
\epsilon_e(l)\;=\; \sum_{E \supset e} \lambda_E \q,\q\q\q
L_E \;=\;\sum_{e\subset E} l_e \q .
\ea
Multiplying the first equation with in (\ref{regge3}) and summing over all edges we find that (Here we indicate with $l(L),\lambda(L)$ that the lengths $l_e(L)$ and Lagrange multiplier $\lambda_E(L)$ satisfy the equations of motion (\ref{regge3}) and hence depend on the $L_E$.)
\ba\label{regge4}
S'\;:=\;{\sum_e l_e(L) \epsilon_e(l(L))} \;=\; \sum_e l_e \sum_{E \supset e} \lambda_E(L) \; = \; \sum_E L_E  \lambda_E(L) \q .
\ea
Now from the first equation of motion (\ref{regge3}) it actually follows that $\lambda_E$ is the deficit angle at $E$ in the coarse grained triangulation (which just agrees with the deficit angles of all the edges $e$ making up $E$). Hence we indeed just obtain again the original Regge action (\ref{regge1}) as the coarse grained action.

\section{Discussion and outlook}

The purpose of this work was to develop some necessary methods, in order to construct discrete actions, that feature continuum (gauge) symmetries. One motivation is to understand how discrete gravity actions with an exact notion of diffeomorphism symmetry could be obtained. As in general relativity diffeomorphism symmetry is deeply intertwined with the dynamics of the theory, investigation of this problem could shed some light on one of the most important  problems in many quantum gravity approaches, namely to show that general relativity emerges in the large scale limit.

The main idea to construct such discrete actions featuring continuum symmetries, is to `pull back' continuum dynamics to the lattice. This can be done via a coarse graining/renormalization approach. The advantage of this method is that in the long term, the tools developed may also help to derive the large scale limit of discrete quantum gravity theories.

Here we took some initial steps in this program. In particular we formulated how to coarse grain theories with gauge degrees of freedom in a gauge covariant way. This is particularly important regarding discrete gravity approaches, where there is a notion of diffeomorphism symmetry for the linearized theory, which is however broken at higher order. Thus gauge fixing would be inconsistent, at least to higher order. Furthermore we provided a discrete gravity action and a coarse graining map for gravity, which was motivated by Regge calculus and its geometric interpretation. Indeed with this choice, there is an obvious interpretation of the gauge degrees of freedom as translation of the vertices in the background geometry, which is preserved under coarse graining. Moreover for 3D gravity, the discrete action is invariant under coarse graining with the coarse graining map provided.

We have only considered free theories -- one obvious development is to consider the theory to higher order, where in the case of discrete gravity, diffeomorphism invariance is broken. Thus one could verify or falsify, whether one regains with these methods diffeomorphism symmetry at least on the perturbative level.

Numerous other directions for further development are possible, we will just list a few of them:
\begin{itemize}
\item If the theories are not topological, the coarse grained actions will in general be non--local. Here it would be interesting to develop a canonical analysis of such non-local theories on the lattice \cite{ralf}.
For canonical lattice gravity a long--standing problem is, that the algebra of (gauge transformation generating) constraints does not close \cite{lollreview,Loll:1997iw}. One possibility is, that to obtain closure of the constraint algebra, one might have to involve canonical formulations of such non--local theories.

\item In the course of this work, we have seen that the complexity of the coarse graining process might depend very much on the choice of basic variables. Indeed for general relativity, there exists a plethora of different formulations in the continuum \cite{plebanski,Ashtekar:1986yd} but also in the discrete \cite{Barrett, arear, Dittrich:2008va, Bahr:2009qd,Dittrich:2008ar,Freidel:2010aq} based on different kind of variables. Here it could be very fruitful to see, which formulations are most amenable for coarse graining. Since many of the formulations on which spinfoams \cite{spinfoams} are based involve (second class) constraints, an interesting problem is to investigate the behavior of such constraints under coarse graining \cite{toappear}.  This could also shed some light on the problem, whether degenerate configurations, which play an important role in spin foam formulations \cite{Barrett:2009as,Dittrich:2010ey}, will be relevant for the large scale dynamics.

\item A related question is to consider alternative coarse graining maps, to see how these influence the locality of the coarse grained action  \cite{Bietenholz:1999kr, Bell:1974vv}, but also whether these alternative coarse grainings regain (diffeomorphism) symmetries.

\item For the quantum theory, one would not only need a diffeomorphism invariant discrete action, but also a diffeomorphism invariant measure for the discrete theory. Such a measure can likewise be obtained by coarse graining/renormalization of the partition function \cite{steinhaus}. This question can even be considered for linearized gravity, however one has to keep some parameters of the triangulation of the flat background geometry as free variables, in order  to gain some information on the measure, which will be a function of these parameters.

\item The actions and partition functions obtained via coarse graining should be explored, in particular with regard to diffeomorphism invariance. Here an interesting more general question is, whether diffeomorphism invariant actions are necessarily fixed points of some coarse graining process, in particular whether these are connected to discretization (or triangulation) independence. If this is the case, the question arises, how this approach, where triangulation independence is obtained via coarse graining, is related to approaches, where triangulation independence is reached via a sum over triangulations \cite{gft}.

\end{itemize}

\section*{Acknowledgements}
BD would like to thank Carlo Rovelli and James Ryan for fruitful discussions about the nature of diffeomorphism symmetries in discrete theories.

\vspace{1cm}

\appendix

\section{Coarse graining map for the metric variables}\label{appmetric}

Here we want to derive geometrically the coarse graining map for the metric variables $h_{ab}$ used in section \ref{Sec:LinearizedGravity}. This map can be obtained by considering a regular (hyper--) cubical lattice to which we assign as variables the length of the edges. It would however not be sufficient to have only the lengths of the edges in the $d$ coordinate directions to reconstruct the full metric. Rather one has to introduce diagonals $e_{bc}$, that is edges starting from a vertex $x$ and ending at $x+e_b+e_c$ where $e_b,e_c$ are the lattice vectors in direction $b,c=1,\ldots, d$. In three and four dimensions it is also necessary to introduce further edges, namely body diagonals along $e_a+e_b+e_c$ and in four dimensions hyperbody diagonals along $e_1+e_2+e_3+e_4$, to obtain a regular triangulation into simplices. But the lengths of these additional edges are subject to a trivial dynamics in linearirized Regge calculus, that is these variables can be ignored \cite{Rocek}. Indeed as we will see below the lengths of the edges $l_a$ along all the directions $e_a$ together with the lengths of the diagonals $l_{ab}$ along the directions $e_{a}+e_{b}$ are sufficient to find all  the lattice metric components $g_{ab}$.

To this end we just need to consider how the lengths of the edges are computed from the metric $g$
\ba\label{metric1}
l_a^2&=&e_a \cdot g \cdot e_a\q\q\q\q\q\q \;=\;  g_{aa}   \nn\\
l_{ab}^2  &=& (e_a+e_b) \cdot g \cdot (e_a+e_b)\,\, \;=\; g_{aa} + 2 g_{ab} + g_{bb}   \q .
\ea

We will consider perturbations for the length variables and for the metric variables
\ba\label{metric2}
l_a\;=\;l^{(0)}_a(1+ \epsilon \lambda_a)  \q,\q\q  l_{ab}\;=\; l^{(0)}_{ab}(1+ \epsilon \lambda_{ab})  \q, \q\q g_{ab} \;=\; g_{ab}^{(0)}+\epsilon h_{ab}
\ea
where the background values are $l^{(0)}_b=a, l^{(0)}_{bc}=\sqrt{2}a$ and $g_{bc}=a^2 \delta_{bc}$ with $a$ the lattice constant. Using the expansion (\ref{metric2}) in the relations (\ref{metric1}) and keeping only terms to first order in $\epsilon$ we obtain
\ba\label{metric3}
h_{aa}\;=\;2a^2 \lambda_a  \q,\q\q  h_{ab}\;=\;a^2\left(2\lambda_{ab}- \lambda_a - \lambda_b \right)  \q .
\ea

The coarse graining of the length variables is straightforward: the lengths of a coarse grained edge $L_a,L_{ab}$ should be given by the sum of the lengths $l_a,l_{ab}$, respectively, of all the edges forming this new coarse grained edge. Hence we obtain for the coarse graining of the perturbation variables  $\lambda_a,\lambda_{ab}$
\ba\label{metric4}
\Lambda_a (X) \;=\;   \frac{1}{L} \sum_{z=0}^{L-1}  \lambda_a(LX+ze_a) \q, \q\q \Lambda_{ab} (X) \;=\;   \frac{1}{L} \sum_{z=0}^{L-1}  \lambda_{ab}(LX+z(e_a+e_b))   \q .
\ea
Using the relations (\ref{metric3}) on both sides of these equations leads to the following coarse graining map for the perturbative metric variables
\ba\label{metric4b}
H_{aa} &=& \q L \sum_{z=0}^{L-1} h_{aa} (LX+ze_a)   \nn\\
H_{ab}  &=& \q L\sum_{z=0}^{L-1} \left(h_{ab}+ \tfrac{1}{2} h_{aa}+\tfrac{1}{2} h_{bb}\right)(LX+ z(e_a+e_b)) \nn\\
&&-   L\sum_{z=0}^{L-1}  \tfrac{1}{2} h_{aa}(X+ze_a)  -L\sum_{z=0}^{L-1} \tfrac{1}{2} h_{bb}(LX+ ze_b)  \q .
\ea
Defining the coarse graining matrix $B_{abcd}(X,x)$ by
\ba\label{metric5}
H_{ab}(X)&=&\sum_{c,d} \sum_x B_{abcd}(X,x)h_{ab}(x)
\ea
we obtain for its Fourier transformation
\ba\label{metric6}
B_{abcd}(P,p)&:=& \sum_{X,x} e^{-2\pi i\frac{P\dot X}{N'}} \,B_{abcd}(X,x)e^{2\pi i \frac{p\dot x}{N}}
\ea
the following result
\ba\label{metric7}
B_{abcd}(P,p)\;=\;\frac{N^d}{L^{d-1} }\delta^{(N')} (P-p)\, \Bigg(\!\!\!\! \!\!\!\!&&
\delta_{ab}\delta_{cd} \; \; \delta_{ac} \frac{K_a}{k_a}
+
\nn\\
&&(1-\delta_{ab})\delta_{cd} \;\frac{1}{2}\left( \delta_{ac} \left(\frac{K_{ab}}{k_{ab}} -\frac{K_a}{k_a}\right) + \delta_{bc} \left(\frac{K_{ab}}{k_{ab}} -\frac{K_b}{k_b} \right)\right) + \nn\\
&&
 (1-\delta_{ab})(1-\delta_{cd}) \;\frac{1}{2}(\delta_{ac}\delta_{bd}+\delta_{ad}\delta_{bc}) \frac{K_{ab}}{k_{ab}}
\Bigg)  \q ,
\ea
where we abbreviated $k_{ab}=1-e^{\frac{2\pi i}{N}p_a} e^{\frac{2\pi i}{N}p_b} =k_a+k_b-k_a k_b$ and similarly for $K_{ab}$.
Note that there is a non--diagonal part in the coarse graining matrix between the diagonal metric elements $h_{cc}$ and the non--diagonal metric elements $H_{ab},\,a\neq b$.

Finally we will show that the gauge modes of the action (\ref{gravaction}) are related to infinitesimal vertex translations.  Changing the positions of the vertices infinitesimally by the amount $\gamma_{a}(x)$ in the direction $e_a$ at the vertex at $x$ leads to a change in the length variables $\lambda_a,\lambda_{ab}$ by
\ba\label{metric8}
\lambda_a(x) &\mapsto  &\lambda_a(x) + \gamma_a(x+e_a)-\gamma_a(x) \nn\\
\lambda_{ab}(x) &\mapsto & \lambda_{ab}+\tfrac{1}{2}\left(\gamma_a(x+e_a+e_b)- \gamma_a(x)+\gamma_b(x+e_a+e_b)-\gamma_b(x) \right) \q  .
\ea
This gives for the metric variables the gauge transformations
\ba\label{metric7b}
h_{aa}(x) &\mapsto  &h_{aa}(x) + 2a^2\left(\gamma_a(x+e_a)-\gamma_a(x)\right) \nn\\
h_{ab}(x) &\mapsto & h_{ab}(x)+a^2\left(\gamma_a(x+e_a+e_b)- \gamma_a(x+e_a)+\gamma_b(x+e_a+e_b)-\gamma_b(x+e_b) \right) \; .\q\q
\ea
Fourier transformation gives the gauge modes
\ba \label{metric8b}
v_{ab}^c(p)=\delta_a^c\,\,k_b\,\,(1-(1-\delta_{ab})k_a)    +\delta^c_b\,\,k_a\,\,(1-(1-\delta_{ab})k_b)
\ea
which are exactly the longitudinal modes projected on by $\Pi^1$.  Note that the form of the modes is left invariant under the coarse graining map
\ba
\sum_{cd}\sum_p  B_{abef}(P,p) v_{ef}^c(p)\; \sim\; v^c_{ab}(P)\,=\, \delta_a^c\,\,K_b\,\,(1-(1-\delta_{ab})K_a)    +\delta^c_b\,\,K_a\,\,(1-(1-\delta_{ab})K_b)   \, .\q
\ea

\section{Some summations}\label{somesums}

In the case of 1d systems and topological models one can actually perform the sums over the fine grained wave vectors $p$. In this section we will provide explicit expressions for some sums needed in the main text.

The sums involve exponential functions of the form
\be
\omega_a \;:=\;1-k_a\;=\;\exp(  \frac{2\pi i}{N}  P_a  + \frac{2\pi i}{L} r_a)\; =:\;\exp(  ix_a + \frac{2\pi i}{L} r_a)
\ee
which are summed over $r_a=0,\ldots,L-1$. Note that $\omega_a^L=\exp(\frac{2\pi i}{N'}  P_a)=:1-K_a$ does not depend on the summation label $r_a$, hence can be pulled out of the sum. To evaluated the sum we will in all cases use a rewriting of the summands into a geometric series. The basic idea is described by
~\\
~\\
{\bf Case A:}
\ba\label{nov1}
\sum_{r=0}^{L-1}\, \frac{1}{k}
 \;:=\; \sum_{r=0}^{L-1} \,\frac{1}{1-e^{ix+  \frac{2\pi i}{L}r}}
&=& \frac{L}{1-e^{iLx}}
\;=:\;  \frac{L}{K}\q.
\ea
{\bf Proof:}
\ba\label{nov2}
\sum_{r=0}^{L-1}\, \frac{1}{k}  &=& \sum_{r=0}^{L-1}\, \frac{1}{1-\omega}  \nn\\
&=&\frac{1}{1-\omega^L}  \sum_{r=0}^{L-1}\, \frac{1-\omega^L}{1-\omega} \nn\\
&=& \frac{1}{1-\omega^L}  \sum_{r=0}^{L-1}\sum_{s=0}^{L-1} \omega^s \;=\;  \frac{1}{1-\omega^L}  \sum_{r=0}^{L-1}\sum_{s=0}^{L-1}e^{ i s x + \frac{2\pi i}{L} r s} \nn\\
 &=&\frac{1}{1-\omega^L} \sum_{s=0}^{L-1}e^{ i s x } \, L \, \delta^{(L)}(s) \nn\\
 &=& \frac{L}{1-\omega^L}  \;=\; \frac{L}{K} \q ,
\ea
where we used that the $L$-periodic delta function is given by
\be
 \delta^{(L)}(s)=\frac{1}{L}\sum_{r=0}^{L-1} e^{\frac{2\pi i}{L} r s}  \q .
\ee
Similarly we obtain the sum needed for 2d electromagnetism in chapter \ref{Sec:Sub2DElectromagnetism}:
~\\
~\\
{\bf Case B:}
\ba\label{nov3}
\sum_{r=0}^{L-1}\, \frac{1}{k \,\bar{k}}
\;:= \;
\sum_{r=0}^{L-1} \,  \frac{1}{   (1-e^{ix+  \frac{2\pi i}{L}r }  )  (1-e^{-ix-  \frac{2\pi i}{L}r } ) }
&=&  \frac{L^2}{(1-e^{iLx})(1-e^{-iLx})}
\;=:\; \frac{L^2}{K \, \bar{K}} \; . \q\;
\ea
{\bf Proof:}
\ba\label{nov4}
\sum_{r=0}^{L-1}\, \frac{1}{k \bar{k}}  &=&
\frac{1}{(1-\omega^L)(1-\omega^{-L})} \sum_{r=0}^{L-1} \sum_{s,t}^{L-1}
e^{ i (s-t) x + \frac{2\pi i}{L} r (s-t)}    \nn\\
&=&
\frac{1}{(1-\omega^L)(1-\omega^{-L})} \sum_{s,t}^{L-1}
e^{ i (s-t) x }\,L\, \delta^{(L)}(s-t)    \nn\\
&=&\frac{L^2}{(1-\omega^L)(1-\omega^{-L})} \;=\;  \frac{L^2}{K \, \bar{K}} \; . \q\;
\ea

The sums for 3d gravity are more involved. The two cases we need are
~\\
~\\
{\bf Case C:}
\ba\label{nov5}
&& \sum_{r_1,r_2,r_3}^{L-1} \frac{ \omega_1 \omega_2 \omega_3 }{   (1-\omega_1\omega_2)(1-\omega_1\omega_3)(1-\omega_2\omega_3) }  \nn\\  && \q\q\q\q\q\q\q\q=
  \frac{ \omega_1^L \omega_2^L \omega_3^L }{   (1-\omega_1^L\omega_2^L)(1-\omega_1^L \omega_3^L)(1-\omega_2^L\omega_3^L) }   \times
\begin{cases}
 0 &  \mbox{if} \; L\; \mbox{is even} \\
 ~\\
 L^3                              &   \mbox{if}\; L\; \mbox{is odd.}  \q\q\q
\end{cases}
\ea
~\\
{\bf Case D:}
\ba\label{nov6}
&&\sum_{r_1,r_2,r_3}^{L-1} \frac{1}{(1-\omega_a)}\frac{ \omega_1 \omega_2 \omega_3 }{   (1-\omega_1\omega_2)(1-\omega_1\omega_3)(1-\omega_2\omega_3) } \nn\\
 &&\q\q\q\q \q\q\q  =
 \frac{ \omega_1^L \omega_2^L \omega_3^L }{   (1-\omega_1^L\omega_2^L)(1-\omega_1^L \omega_3^L)(1-\omega_2^L\omega_3^L) }      \times
 \begin{cases}
  \frac{L^4}{2} \frac{1+\omega_a^L}{1-\omega_a^L}     \q\q \;
 &  \mbox{if} \; L\; \mbox{is even} \\
 ~\\
  \frac{L^4}{2}    \frac{1+\omega_a^L}{1-\omega_a^L}
  +  \frac{L^3}{2}         &   \mbox{if}\; L\; \mbox{is odd.}  \q\q\q
\end{cases}
\ea
~\\
{\bf Proof:}\\
Let us first consider the sum over $r_1$ for Case C. To this end we write the $\omega_1$ dependent part as
\ba\label{nov7}
\frac{\omega_1}{(1-\omega_1 \omega_2)(1-\omega_1 \omega_3)}&=&  \frac{-\omega^{-1}_3}{  (1-\omega_1 \omega_2)(1-\omega_1^{-1} \omega_3^{-1})}  \q .
\ea
Using again the rewriting of these terms into a geometric series we arrive at
\ba\label{nov8}
\sum_{r_1=0}^{L-1} \frac{\omega_1}{(1-\omega_1 \omega_2)(1-\omega_1 \omega_3)}
&=&
\frac{-\omega_3^{-1}} {(1-\omega_1^L \omega_2^L)(1-\omega_1^{-L} \omega_3^{-L})}
\sum_{s,t=0}^{L-1} \sum_{r_1=0}^{L-1}   e^{ i x_1 (s-t) + \frac{2\pi i}{L} r_1 (s-t)} \, \omega_2^s \omega_3^{-t} \q \nn\\
&=&\frac{-\omega_3^{-1}} {(1-\omega_1^L \omega_2^L)(1-\omega_1^{-L} \omega_3^{-L})}
\sum_{s,t=0}^{L-1}  L\,\delta^{(L)}\!(s-t)\,\,  e^{ i x_1 (s-t) }\,\, \omega_2^s \omega_3^{-t} \q \nn\\
&=&\frac{-L \,\omega_3^{-1}} {(1-\omega_1^L \omega_2^L)(1-\omega_1^{-L} \omega_3^{-L})}
\sum_{s=0}^{L-1}  \left( \omega_2 \omega_3^{-1}  \right)^s \nn\\
&=&\frac{-L \,\omega_3^{-1}} {(1-\omega_1^L \omega_2^L)(1-\omega_1^{-L} \omega_3^{-L})}
\frac{(1-  \omega_2^L \omega_3^{-L})}  {(  1 -   \omega_2 \omega_3^{-1})  }   \q .
\ea
This leaves us with the following $\omega_2$ dependent terms  from (\ref{nov5}) and (\ref{nov8})
\ba\label{nov9}
\frac{\omega_2}{(1-\omega_2\omega_3)} \frac{1}{(  1 -   \omega_2 \omega_3^{-1})  }
&=&\frac{-\omega_3}{ (1-\omega_2\omega_3)} \frac{1}{(  1 -   \omega_2^{-1} \omega_3)   }  \q .
\ea
A similar calculation as in (\ref{nov8}) leads to
\ba\label{nov10}
\sum_{r_2=0}^{L-1} \frac{\omega_2}{(1-\omega_2\omega_3)} \frac{1}{(  1 -   \omega_2 \omega_3^{-1})  }
&=&
\frac{-L \omega_3}{ (1-\omega_2^L\omega_3^L) (  1 -   \omega_2^{-L} \omega_3^L)   } \frac{(1-\omega_3^{2L})}{(1-\omega_3^2)}  \q .
\ea
Now the remaining $\omega_3$ dependent terms  from (\ref{nov5},\ref{nov8}) and (\ref{nov10})  are given by
\ba\label{nov11}
\frac{\omega_3}{(1-\omega_3^2)} &=&\frac{1}{(\omega^{-1}_3-\omega_3)} \nn\\ &=& \frac{1}{(1-\omega_3)}-\frac{1}{(1-\omega_3^2)}    \q .
\ea
From the symmetric form of the second expression in (\ref{nov11}) one can conclude that the sum over $r_3$ vanishes for even L (as in this case there are $\tfrac{L}{2}$ terms differing by a minus sign from the other $\tfrac{L}{2}$ terms). But we will also find this by evaluating separately the two summands in the second line of (\ref{nov11}). For the first term apply Case A:
\ba\label{nov12}
\sum_{r_3=0}^{L-1}  \frac{1}{(1-\omega_3)} &=& \frac{L}{(1-\omega_3^L)} \q .
\ea
For the second term we obtain
\ba\label{nov13}
\sum_{r_3=0}^{L-1}  \frac{1}{(1-\omega_3^2)}
&=&
  \frac{1}{(1-\omega_3^{2L})}  \sum_{s=0}^{L-1} L \, \delta^{(L)}(2s) e^{i 2x_3 s}  \nn\\
  &=&
    \frac{L}{(1-\omega_3^{2L})}
  \begin{cases}
   (1+ \omega^L_3) &  \mbox{if} \; L\; \mbox{is even} \\
1                          &   \mbox{if}\; L\; \mbox{is odd,}  \q\q\q
\end{cases}
\ea
as for even $L$ there are two solutions to the $L$--periodic delta function $s=0$ and $s=\tfrac{L}{2}$.

Summing up the two contributions in (\ref{nov12}) and (\ref{nov12}) we obtain
\ba\label{nov14}
\sum_{r_3=0}^{L-1}  \frac{\omega_3}{(1-\omega_3^2)}
  &=&
    \frac{L\,\omega_3^L}{(1-\omega_3^{2L})}
  \begin{cases}
   0 &  \mbox{if} \; L\; \mbox{is even} \\
1                          &   \mbox{if}\; L\; \mbox{is odd.}  \q\q\q
\end{cases}
\ea

We have finally performed all the sums for Case C. Collecting all the results of (\ref{nov8},\ref{nov10}) and (\ref{nov14}) we indeed arrive at the claim in (\ref{nov5}).

To tackle Case D it is due to symmetry sufficient to consider the case $\omega_a=\omega_3$. Hence we can re-use most of the calculations for Case C and just redo the summation over the $\omega_3$--dependent terms. These are now given by
\ba\label{nov15}
\frac{\omega_3}{(1-\omega_3)(1-\omega_3^2)}  &=& \frac{1}{(1-\omega_3)^2}-\frac{1}{(1-\omega_3)(1-\omega_3^2)}    \q .
\ea
For the first term we calculate
\ba\label{nov16}
\sum_{r_3=0}^{L-1}  \frac{1}{(1-\omega_3)^2}
  &=&
\frac{1}{(1-\omega_3^L)^2} \sum_{r_3=0}^{L-1}   \sum_{s,t=0}^{L-1} \omega_3^{s+t}  \nn\\
&=&
\frac{L}{(1-\omega_3^L)^2}      \sum_{s,t=0}^{L-1}  e^{ix_3(s+t)} \, \delta^{(L)}\!(s+t)  \nn\\
  &=&
  \frac{L}{(1-\omega_3^L)^2}  \left[1+(L-1)\omega_3^L\right]  \q ,
\ea
as there are two kind of solutions to the $L$--periodic delta function: one is $s,t=0$ and another $L-1$ solutions are given by $s+t=L$.
The second term gives similarly
\ba\label{nov17}
\sum_{r_3=0}^{L-1}  \frac{1}{(1-\omega_3)(1-\omega^2_3)}
  &=&
\frac{1}{(1-\omega_3^L)(1-\omega_3^{2L})} \sum_{r_3=0}^{L-1}   \sum_{s,t=0}^{L-1} \omega_3^{s+2t}  \nn\\
&=&\frac{L}{(1-\omega_3^L)(1-\omega_3^{2L})}   \sum_{s,t=0}^{L-1}   e^{ix_3(s+2t)}  \, \delta^{(L)}\!(s+2t)  \nn\\
&=&
\frac{L}{(1-\omega_3^L)(1-\omega^{2L}_3)} \times
\begin{cases}
1+  \tfrac{L}{2}\omega_3^L +\tfrac{L-2}{2}\omega^{2L}_3 &  \mbox{if} \; L\; \mbox{is even} \\
1+\tfrac{L-1}{2}  \omega_3^L +\tfrac{L-1}{2}\omega^{2L}_3                    &   \mbox{if}\; L\; \mbox{is odd.} \q\q \q
\end{cases}
\ea
Again, collecting the results of (\ref{nov8},\ref{nov10}) and (\ref{nov16},\ref{nov17}) we  arrive at the claim (\ref{nov6}) for Case D.



\begin{thebibliography}{99}
\parskip -2pt




  \bibitem{lollreview}
R.~Loll,
``Discrete Approaches to Quantum Gravity in Four Dimensions'',
Living Reviews in Relativity 13 (1998) online article: [http://relativity.livingreviews.org/Articles/lrr-1998-13/].


\bibitem{Loll:1997iw}
  R.~Loll,
  ``On the diffeomorphism-commutators of lattice quantum gravity,''
  Class.\ Quant.\ Grav.\  {\bf 15} (1998) 799
  [arXiv:gr-qc/9708025].

\bibitem{Dittrich:2008pw}
  B.~Dittrich,
  ``Diffeomorphism symmetry in quantum gravity models,''  Adv.\ Sci.\ Lett.\ {\bf 2} (2009) 151,
  arXiv:0810.3594 [gr-qc].

\bibitem{bahrdittrich1}
 B.~Bahr and B.~Dittrich,
  ``(Broken) Gauge Symmetries and Constraints in Regge Calculus,''
  Class.\ Quant.\ Grav.\  {\bf 26} (2009) 225011
  [arXiv:0905.1670 [gr-qc]].\\
  ``Breaking and restoring of diffeomorphism symmetry in discrete gravity,''
THE PLANCK SCALE: Proceedings of the XXV Max Born Symposium, Wroclaw (Poland), 29 June - 3 July, 2009, edited by J.~Kowalski-Glikman et.\ al.\ ,pp. 10-17
  arXiv:0909.5688 [gr-qc].





\bibitem{Dittrich:2009fb}
  B.~Dittrich and P.~A.~Hohn,
  ``From covariant to canonical formulations of discrete gravity,''
  Class.\ Quant.\ Grav.\  {\bf 27} (2010) 155001
  [arXiv:0912.1817 [gr-qc]].



\bibitem{Thiemann:2003zv}
  T.~Thiemann,
  ``The Phoenix project: Master constraint programme for loop quantum
  gravity,''
  Class.\ Quant.\ Grav.\  {\bf 23} (2006) 2211
  [arXiv:gr-qc/0305080].\\
   B.~Dittrich and T.~Thiemann,
  ``Testing the master constraint programme for loop quantum gravity. I:
  General framework,''
  Class.\ Quant.\ Grav.\  {\bf 23} (2006) 1025
  [arXiv:gr-qc/0411138].\\
 K.~Giesel and T.~Thiemann,
  ``Algebraic Quantum Gravity (AQG) I. Conceptual Setup,''
  Class.\ Quant.\ Grav.\  {\bf 24} (2007) 2465
  [arXiv:gr-qc/0607099].







\bibitem{Gambini:2005jm}
  R.~Gambini and J.~Pullin,
  ``Consistent discretizations as a road to quantum gravity,''
  arXiv:gr-qc/0512065.\\
  M.~Campiglia, C.~Di Bartolo, R.~Gambini and J.~Pullin,
  ``Uniform discretizations: A quantization procedure for totally  constrained
  systems including gravity,''
  J.\ Phys.\ Conf.\ Ser.\  {\bf 67} (2007) 012020
  [arXiv:gr-qc/0606121].\\
 M.~Campiglia, C.~Di Bartolo, R.~Gambini and J.~Pullin,
  ``Uniform discretizations: A new approach for the quantization of totally
  constrained systems,''
  Phys.\ Rev.\  D {\bf 74} (2006) 124012
  [arXiv:gr-qc/0610023].\\
 R.~Gambini and J.~Pullin,
  ``Emergent diffeomorphism invariance in a discrete loop quantum gravity
  model,''
  Class.\ Quant.\ Grav.\  {\bf 26} (2009) 035002
  [arXiv:0807.2808 [gr-qc]].

\bibitem{wp}
  T.~Piran and R.~M.~Williams,
  ``A (3+1) Formulation Of Regge Calculus,''
  Phys.\ Rev.\  D {\bf 33} (1986) 1622.

\bibitem{friedman}
 J.~L.~Friedman and I.~Jack,
  ``(3+1) Regge Calculus With Conserved Momentum And Hamiltonian Constraints,''
  J.\ Math.\ Phys.\  {\bf 27} (1986) 2973.


\bibitem{miller}
W.~A.~Miller,``The geometrodynamic content of the Regge equations as illuminated by the boundary of a boundary principle,'' Found.\ Phys.\ {\bf 16} (1986) 143.
 A.~P.~Gentle, A.~Kheyfets, J.~R.~McDonald and W.~A.~Miller,
  ``A Kirchoff-like conservation law in Regge calculus,''
  arXiv:0807.3041 [gr-qc].


\bibitem{morse} 
P.~A.~Morse,
  ``Approximate diffeomorphism invariance in near flat simplicial geometries,''
Class.\ Quant.\ Grav.\ {\bf 9} (1992) 2489.






\bibitem{hwgauge}
H.~W.~Hamber and R.~M.~Williams,
  ``Gauge invariance in simplicial gravity,''
  Nucl.\ Phys.\  B {\bf 487} (1997) 345
  [arXiv:hep-th/9607153].





\bibitem{menotti}
  P.~Menotti and P.~P.~Peirano,
  ``Diffeomorphism invariant measure for finite dimensional geometries,''
  Nucl.\ Phys.\  B {\bf 488} (1997) 719
  [arXiv:hep-th/9607071].





\bibitem{Hasenfratz:1997ft}
P.~Hasenfratz and F.~Niedermayer,
  ``Perfect Lattice Action For Asymptotically Free Theories,''
  Nucl.\ Phys.\  B {\bf 414} (1994) 785
  [arXiv:hep-lat/9308004].\\
  P.~Hasenfratz,
  ``Prospects for perfect actions,''
  Nucl.\ Phys.\ Proc.\ Suppl.\  {\bf 63} (1998) 53
  [arXiv:hep-lat/9709110].



\bibitem{Bahr:2009qc}
  B.~Bahr and B.~Dittrich,
  ``Improved and Perfect Actions in Discrete Gravity,''
  Phys.\ Rev.\  D {\bf 80} (2009) 124030
  [arXiv:0907.4323 [gr-qc]].

\bibitem{Bietenholz:1999kr}
  W.~Bietenholz,
  ``Perfect scalars on the lattice,''
  Int.\ J.\ Mod.\ Phys.\  A {\bf 15} (2000) 3341
  [arXiv:hep-lat/9911015].





\bibitem{Wilson:1973jj}
  K.~G.~Wilson and J.~B.~Kogut,
  ``The Renormalization group and the epsilon expansion,''
  Phys.\ Rept.\  {\bf 12} (1974) 75.

\bibitem{Bell:1974vv}
  T.~L.~Bell and K.~G.~Wilson,
  ``Nonlinear Renormalization Groups,''
  Phys.\ Rev.\ B {\bf 10} (1974) 3935.\\
 `` Finite-lattice approximations to renormalization groups"
 Phys.\ Rev.\ B {\bf 11} (1975) 3431

\bibitem{Bietenholz:1995cy}
  W.~Bietenholz and U.~J.~Wiese,
  ``Perfect Lattice Actions for Quarks and Gluons,''
  Nucl.\ Phys.\  B {\bf 464} (1996) 319
  [arXiv:hep-lat/9510026].


\bibitem{Regge}
  T.~Regge,
  ``General relativity without coordinates,''
  Nuovo Cim.\  {\bf 19} (1961) 558.\\
  T.~Regge and R.~M.~Williams,
  ``Discrete structures in gravity,''
  J.\ Math.\ Phys.\  {\bf 41} (2000) 3964
  [arXiv:gr-qc/0012035].\\
R.~M.~Williams and P.~A.~Tuckey,
``Regge calculus: a brief review and bibliography"
Class.\ Quant.\ Grav.\ {\bf 9} (1992) 1409



\bibitem{Bahr:2009qd}
  B.~Bahr and B.~Dittrich,
  ``Regge calculus from a new angle,''
  New J.\ Phys.\  {\bf 12} (2010) 033010
  [arXiv:0907.4325 [gr-qc]].

  \bibitem{Rocek}
  M.~Rocek and R.~M.~Williams,
  ``Quantum Regge Calculus,''
  Phys.\ Lett.\ B {\bf 104} (1981) 31.
\\
  ``The Quantization Of Regge Calculus,''
  Z.\ Phys.\ C {\bf 21} (1984) 371.
\\
H.~W.~Hamber and R.~M.~Williams,
  ``Simplicial quantum gravity in three-dimensions: Analytical and numerical results,''
  Phys.\ Rev.\  D {\bf 47} (1993) 510.
  \\
  ``Non-perturbative gravity and the spin of the lattice graviton,''
  Phys.\ Rev.\  D {\bf 70} (2004) 124007
  [arXiv:hep-th/0407039].\\
%
  B.~Dittrich, L.~Freidel and S.~Speziale,
  ``Linearized dynamics from the 4-simplex Regge action,''
  Phys.\ Rev.\  D {\bf 76} (2007) 104020
  [arXiv:0707.4513 [gr-qc]].

\bibitem{Batrouni:1984rb}
  G.~G.~Batrouni,
  ``Plaquette formulation and the Bianchi identity for lattice gauge theory,''
  Nucl.\ Phys.\  B {\bf 208} (1982) 467.


\bibitem{toappear}
B.~Dittrich,
``Coarse graining with constraints", to appear


\bibitem{barrett}
J.~W.~Barrett,
``A convergence result for linearized Regge calculus,"
Class.\ Quant.\ Grav.\ {\bf 5} (1988) 1187\\
J.~W.~Barrett, R.~M.~Williams,
``The convergence of lattice solutions of linearised
Regge calculus,"
Class.\ Quant.\ Grav.\ {\bf 5} (1988) 1543


\bibitem{ralf}
R.~Banisch and B.~Dittrich,
``Canonical formulation of non--local lattice theories", to appear



\bibitem{plebanski}
 J.~F.~Plebanski,
  ``On the separation of Einsteinian substructures,''
  J.\ Math.\ Phys.\  {\bf 18} (1977) 2511.




\bibitem{Ashtekar:1986yd}
  A.~Ashtekar,
  ``New Variables for Classical and Quantum Gravity,''
  Phys.\ Rev.\ Lett.\  {\bf 57} (1986) 2244.





\bibitem{Barrett}
  J.~W.~Barrett,
  ``First order Regge calculus,''
  Class.\ Quant.\ Grav.\  {\bf 11}, 2723 (1994)
  [arXiv:hep-th/9404124].

\bibitem{arear}
  J.~W.~Barrett, M.~Rocek and R.~M.~Williams,
  ``A note on area variables in Regge calculus,''
  Class.\ Quant.\ Grav.\  {\bf 16}, 1373 (1999)
  [arXiv:gr-qc/9710056].\\
%
  C.~Wainwright and R.~M.~Williams,
  ``Area Regge calculus and discontinuous metrics,''
  Class.\ Quant.\ Grav.\  {\bf 21} (2004) 4865
  [arXiv:gr-qc/0405031].



\bibitem{Dittrich:2008va}
  B.~Dittrich and S.~Speziale,
  ``Area-angle variables for general relativity,''
  New J.\ Phys.\  {\bf 10} (2008) 083006
  [arXiv:0802.0864 [gr-qc]].



  \bibitem{Dittrich:2008ar}
  B.~Dittrich and J.~P.~Ryan,
  ``Phase space descriptions for simplicial 4d geometries,''
  arXiv:0807.2806 [gr-qc].


\bibitem{Freidel:2010aq}
  L.~Freidel and S.~Speziale,
  ``Twisted geometries: A geometric parametrisation of SU(2) phase space,''
  Phys.\ Rev.\  D {\bf 82} (2010) 084040
  [arXiv:1001.2748 [gr-qc]].








\bibitem{spinfoams}
A.~Perez,
  ``Spin foam models for quantum gravity,''
  Class.\ Quant.\ Grav.\  {\bf 20} (2003) R43
  [arXiv:gr-qc/0301113]. \\
  ``Introduction to loop quantum gravity and spin foams,''
  arXiv:gr-qc/0409061.




\bibitem{Barrett:2009as}
  J.~W.~Barrett, W.~J.~Fairbairn and F.~Hellmann,
  ``Quantum gravity asymptotics from the SU(2) 15j symbol,''
  Int.\ J.\ Mod.\ Phys.\  A {\bf 25} (2010) 2897
  [arXiv:0912.4907 [gr-qc]].



\bibitem{Dittrich:2010ey}
  B.~Dittrich and J.~P.~Ryan,
  ``Simplicity in simplicial phase space,''
  Phys.\ Rev.\  D {\bf 82} (2010) 064026
  [arXiv:1006.4295 [gr-qc]].






\bibitem{steinhaus}
B.~Bahr, B.~Dittrich and S.~Steinhaus,
  ``Perfect discretization of reparametrization invariant path integrals,''
  arXiv:1101.4775 [gr-qc].

\bibitem{gft}
    L.~Freidel,
   ``Group field theory: An overview,''
  Int.\ J.\ Theor.\ Phys.\  {\bf 44} (2005) 1769
  [arXiv:hep-th/0505016].\\
D.~Oriti,
  ``The group field theory approach to quantum gravity,''
  arXiv:gr-qc/0607032.



\end{thebibliography}
\end{document}